\documentclass[fleqn,usenatbib]{mnras}

\usepackage{newtxtext,newtxmath}
\usepackage[T1]{fontenc}

\DeclareRobustCommand{\VAN}[3]{#2}
\let\VANthebibliography\thebibliography
\def\thebibliography{\DeclareRobustCommand{\VAN}[3]{##3}\VANthebibliography}

\usepackage{graphicx}	% Including figure files
\usepackage{amsmath}	% Advanced maths commands
\usepackage{siunitx}
\usepackage{pifont}% http://ctan.org/pkg/pifont
\newcommand{\angstrom}{\text{\normalfont\AA}}
\newcommand{\fesc}{$f_{\rm esc}$}

\title[Two Modes of LyC Escape]{Two Modes of LyC Escape From Bursty Star Formation: Implications for [C~{\small II}] Deficits and the Sources of Reionization}

\author[H. Katz] {Harley Katz$^{1}$\thanks{E-mail:
  \href{mailto:harley.katz@physics.ox.ac.uk}{harley.katz@physics.ox.ac.uk}}, Aayush Saxena$^2$, Joki Rosdahl$^3$, Taysun Kimm$^4$, Jeremy Blaizot$^3$, Thibault Garel$^5$, \newauthor Leo Michel-Dansac$^3$, Martin Haehnelt$^6$, Richard S. Ellis$^2$, Laura Penterrici$^7$, Julien Devriendt$^1$, \newauthor \& Adrianne Slyz$^1$
  \\
  $^1$Sub-department of Astrophysics, University of Oxford,
   Keble Road, Oxford OX1 3RH, United Kingdom \\
  $^2$Department of Physics and Astronomy, University College London, Gower Street, London WC1E 6BT, United Kingdom \\
  $^3$Univ Lyon, Univ Lyon1, Ens de Lyon, CNRS, Centre de Recherche Astrophysique de Lyon UMR5574, F-69230, Saint-Genis-Laval, France \\
  $^4$Department of Astronomy, Yonsei University, 50 Yonsei-ro,
  Seodaemun-gu, Seoul 03722, Republic of Korea \\  
  $^5$Observatoire de Genève, Université de Genève, Chemin Pegasi 51, 1290 Versoix, Switzerland\\
  $^6$Kavli Institute for Cosmology and Institute of Astronomy, Madingley Road, Cambridge CB3 0HA, United Kingdom\\
  $^7$INAF - Osservatorio Astronomico di Roma, via Frascati 33, 00078, Monteporzio Catone, Italy
  }

\date{Accepted XXX. Received YYY; in original form ZZZ}

\pubyear{2022}

\begin{document}
\label{firstpage}
\pagerange{\pageref{firstpage}--\pageref{lastpage}}
\maketitle

\begin{abstract}
We use the SPHINX$^{20}$ cosmological radiation hydrodynamics simulation to study how Lyman Continuum (LyC) photons escape from galaxies and the observational signatures of this escape. We define two classes of LyC leaker: Bursty Leakers and Remnant Leakers, based on their star formation rates (SFRs) that are averaged over 10~Myr (SFR$_{10}$) or 100~Myr (SFR$_{100}$). Both have $f_{\rm esc}>20\%$ and experienced an extreme burst of star formation, but Bursty Leakers have ${\rm SFR_{10}>SFR_{100}}$, while Remnant Leakers have ${\rm SFR_{10}<SFR_{100}}$. The maximum SFRs in these bursts were typically $\sim100$ times greater than the SFR of the galaxy prior to the burst, a rare $2\sigma$ outlier among the general high-redshift galaxy population. Bursty Leakers are qualitatively similar to ionization-bounded nebulae with holes, exhibiting high ionization parameters and typical H{\small II} region gas densities. Remnant Leakers show properties of density-bounded nebulae, having normal ionization parameters but much lower H{\small II} region densities. Both types of leaker exhibit [C{\small II}]$_{\rm 158\mu m}$ deficits on the [C{\small II}]-SFR$_{100}$ relation, while only Bursty Leakers show deficits when SFR$_{10}$ is used. We predict that [C{\small II}] luminosity and SFR indicators such as H$\alpha$ and M$_{\rm 1500\angstrom}$ can be combined to identify both types of LyC leaker and the mode by which photons are escaping. These predictions can be tested with [C{\small II}] observations of known $z=3-4$ LyC leakers. Finally, we show that leakers with $f_{\rm esc}>20\%$ dominate the ionizing photon budget at $z\gtrsim7.5$ but the contribution from galaxies with $f_{\rm esc}<5\%$ becomes significant at the tail-end of reionization.
\end{abstract}

\begin{keywords}
 galaxies: evolution, galaxies: formation, galaxies: high-redshift, stars: formation, ISM: evolution, ISM: general 
\end{keywords}

%%%%%%%%%%%%%%%%%%%%%%%%%%%%%%%%%%%%%%%%%%%%%%%%%%

%%%%%%%%%%%%%%%%% BODY OF PAPER %%%%%%%%%%%%%%%%%%

\section{Introduction}
While it is well established that the Universe completed reionization sometime in the redshift interval of $z=5-7$ \citep{Fan2006,Kulkarni2019}, with Ly$\alpha$ forest data suggesting islands of neutral gas extending to $z\sim5.3$ \citep{Bosman2022}, uncertainties remain on the timing of the onset of reionization, the neutral fraction history, and the sources responsible for the ionizing photons. Empirical constraints on all three are important for understanding the formation of the first stars, metal and dust production in the early Universe, the impact of an emerging UV background on galaxy formation, the visibility of various emission lines (e.g. Ly$\alpha$) at high-redshift, as well as many other characteristics of galaxy formation at cosmic dawn.

Analytic models of reionization that rely on a star formation rate density, an ionizing photon emissivity per unit star formation ($\xi_{\rm ion}$), and a Lyman Continuum (LyC) escape fraction ($f_{\rm esc}$) are often used \citep[e.g.][]{Madau1999,Robertson2013} to model the evolution of the cosmic neutral fraction and constrain the reionization history. While the star formation rate density as a function of redshift can be constrained with observations of the UV luminosity function \citep[e.g.][]{Bouwens2021}, and estimates on $\xi_{\rm ion}$ can be adopted from theoretical stellar evolution models \citep[e.g.][]{Leitherer1999} or inferred from observations \citep[e.g.][]{Stark2015,Bouwens2016}, the majority of the uncertainty in these models stems from our inability to constrain $f_{\rm esc}$. This is due to the fact that $f_{\rm esc}$ cannot be calculated analytically as it is subject to the detailed properties of the interstellar medium (ISM) and the distribution of the sources within. 

Because of the intervening intergalactic medium (IGM), it is nearly impossible to directly observe escaping LyC radiation during the epoch of reionization. Rather observational studies of $f_{\rm esc}$ often target lower redshift analogues of high-redshift galaxies at $z\sim0$ \citep[e.g.][]{Flury2022}, $z\sim3$ \citep[e.g.][]{Fletcher2019}, and more recently at $z\sim1.5$ \citep[e.g.][]{Saha2020}. The limited numbers of observed LyC leakers has historically inhibited a detailed study of their galaxy properties in relation to the general galaxy population. Recent large scale surveys are now making this possible \citep[e.g.][]{Flury2022}; however, the total number of confirmed LyC leakers is still only $\sim100$. Thus any biases in selection function and low number statistics may still contribute significantly to any observed trends between galaxy properties and LyC leakage. For this reason, numerical simulations that resolve the ISM of galaxies remain an invaluable tool for understanding the physics that controls LyC leakage and the relation to galaxy properties \citep[e.g.][]{Katz2020}.

\cite{Zackrisson2013} envisioned two mechanisms by which LyC photons escape from galaxies: a radiation bounded nebula with holes, where the escape fraction is set by the covering fraction of the holes, and a density bounded nebula, where the escape fraction is set by the optical depth of the nebula. These scenarios are not necessarily mutually exclusive; nevertheless, they demonstrate that there exist certain observational signatures that differentiate these mechanisms, especially when dust is included. High-resolution cosmological simulations seem to indicate that $f_{\rm esc}$ is a feedback-regulated quantity \citep[e.g.][]{Trebitsch2017,Kimm2017,Rosdahl2018,Barrow2020}. Bursts of star formation that generate a large quantity of LyC photons can heat and reduce the density of the ISM, occasionally creating holes in the gas distribution where LyC photons escape. This process is then followed by supernova (SN) feedback that can clear even larger channels or destroy the structure of the ISM entirely. Hence, numerical simulations also predict two modes of $f_{\rm esc}$. In the first scenario, early stellar feedback in the form of ionizing radiation and perhaps SNe from the most massive stars creates the first channels through which LyC photons can escape. Because the stellar populations are still young, the LyC production efficiency remains high \citep[e.g.][]{Eldridge2008,Stanway2016}. In the second scenario, later stellar feedback in the form of SNe create super bubbles, clearing out further channels for LyC photons to escape. However, because the LyC production efficiency drops significantly as a function of the age of the stellar population, it is not clear which phase results in more LyC photons leaking into the IGM. Once again, these two modes are not mutually exclusive and it is often the case that one is followed by the other \citep[e.g.][]{Wise2009,Kimm2014,Trebitsch2017,Kimm2017,Rosdahl2018}. \cite{Rosdahl2018} demonstrate this in their Figure~12, where, for a particular galaxy, they show a 50~Myr time series of the evolution of $f_{\rm esc}$, the LyC luminosity, and the structure of the ISM after a strong burst of star formation. The two modes of $f_{\rm esc}$ can also be seen sequentially in the multi-peaked distribution of $f_{\rm esc}$ over time in Figure~8 of \cite{Kimm2017}.

While numerical simulations have predicted the mechanisms by which LyC photons escape galaxies during reionization, they have also demonstrated that the process is inefficient --- at any given time, only a very small fraction of galaxies exhibit high $f_{\rm esc}$ \citep[e.g.][]{Paardekooper2015} and only a few percent of the ionizing photons produced by galaxies escape into the IGM. Hence it is not surprising that observers struggle with finding large populations of LyC leakers \citep[e.g.][]{Leitet2013}. Similarly, simulations also show that the escape fraction is highly viewing angle-dependent \citep[e.g.][]{Cen2015}. Thus, even if a galaxy is a LyC leaker, the probability of directly observing LyC photons can significantly decrease due to geometrical effects.

Of the numerous methods that have been suggested as indirect tracers of LyC leakage, the vast majority rely on emission lines at UV or optical wavelengths. These include high O$_{32}$ ratios \citep[e.g.][]{Izotov2018}, Ly$\alpha$ peak separations \citep[e.g.][]{Verhamme2015,Verhamme2017}, S{\small II} deficits \citep{Wang2019}, Mg{\small II} doublet flux ratios \citep{Chisholm2020}, strong C\,\textsc{iv} emission \citep{Schaerer2022, Saxena2022b}, the combination of H$\beta$ and UV slope \citep{Zackrisson2013}, etc. Because such lines are subject to dust attenuation, and in the case of resonant lines, the gas distribution of the relevant species, line strengths and shapes might also be highly orientation angle-dependent. In contrast, IR emission lines are not as sensitive to geometrical effects. \cite{Katz2020} explored the use of the [C{\small II}]$_{\rm 158\mu m}$ and [O{\small III}]$_{\rm 88\mu m}$ lines as probes of $f_{\rm esc}$ because numerical simulations show that at high-redshift [C{\small II}] emission correlates with the presence of neutral gas \citep{Pallottini2017,Katz2019,Lupi2020} while [O{\small III}] emission traces star formation and feedback \citep[e.g.][]{Katz2019,Lupi2020}. Thus the ratio of the two emission lines is expected to provide insight into $f_{\rm esc}$ \citep{Inoue2016,Katz2020,Katz2021b}. While there are now more than 100 galaxies at high-redshift with [C{\small II}] observations \citep[e.g.][]{LeFevre2020,Bouwens2021}, observations of [O{\small III}] at $z\gtrsim4$ remain limited \citep[e.g.][]{Carniani2020}. The lack of known [O{\small III}] emitters motivates the study of the prospect of using only [C{\small II}] as a means of both identifying potential leakers and the mechanisms by which LyC photons escape. Since the vast majority of known [C{\small II}] emitters are at $z<6$, they can, in principle, be followed up with direct observations in the LyC bands (although IGM transmission can be a problem even at intermediate redshifts \citep{Inoue2008}). Furthermore, with the recent launch of the James Webb Space Telescope (JWST), there is potential to follow up many of the galaxies observed in [C{\small II}] with observations in the rest-frame UV and optical. For this reason, in this work, we use cosmological simulations to study how [C{\small II}] combined with various star formation rate indicators at intermediate and high redshift can be used to identify LyC leakers and differentiate the mechanisms by which LyC photons escape.

This work is organized as follows. In Section~\ref{methods} we briefly review the SPHINX$^{20}$ simulation that resolves $3\times10^4$ galaxies at $z\leq6$. In Section~\ref{results} we discuss the utility of [C{\small II}] as an $f_{\rm esc}$ indicator and highlight synergies between ALMA and other observational facilities that can be used to constrain the properties of the high-redshift ISM. Finally, in Section~\ref{conclusions}, we present our discussion and conclusions.

\begin{figure}
\centerline{\includegraphics[scale=1,trim={0 0.0cm 0cm 0cm},clip]{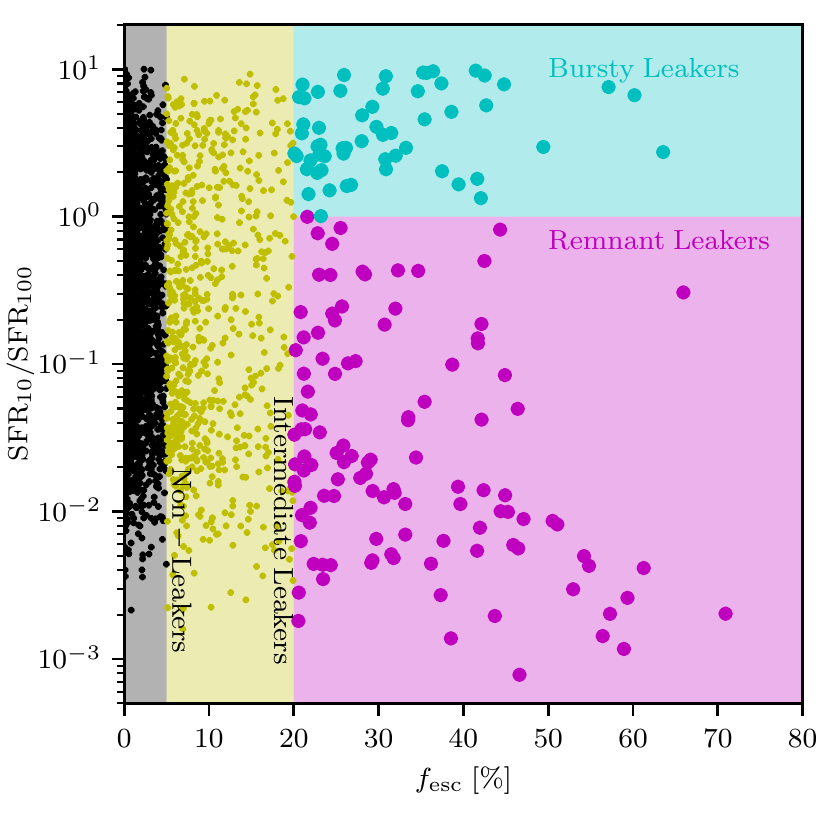}}
\caption{Ratio of 10~Myr-averaged SFR to 100~Myr-averaged SFR versus LyC escape fraction for SPHINX$^{20}$ galaxies at $z=4.64$. The different coloured regions represent the different classifications, as labelled on the diagram.}
\label{classify}
\end{figure}

\begin{figure*}
\centerline{
\includegraphics[scale=1,trim={0 0.0cm 0cm 0cm},clip]{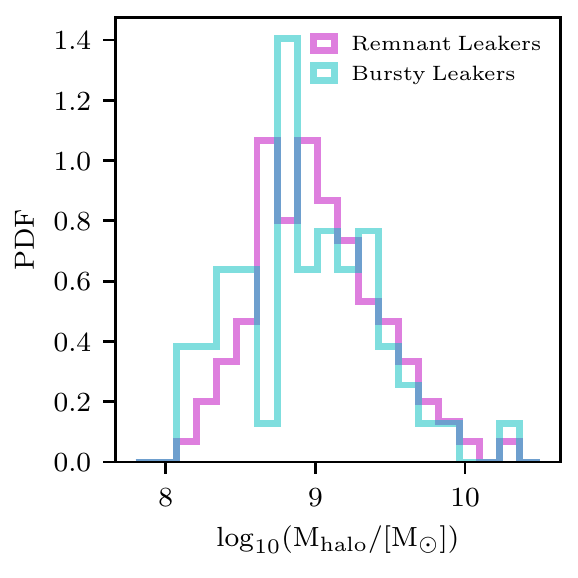}
\includegraphics[scale=1,trim={0 0.0cm 0cm 0cm},clip]{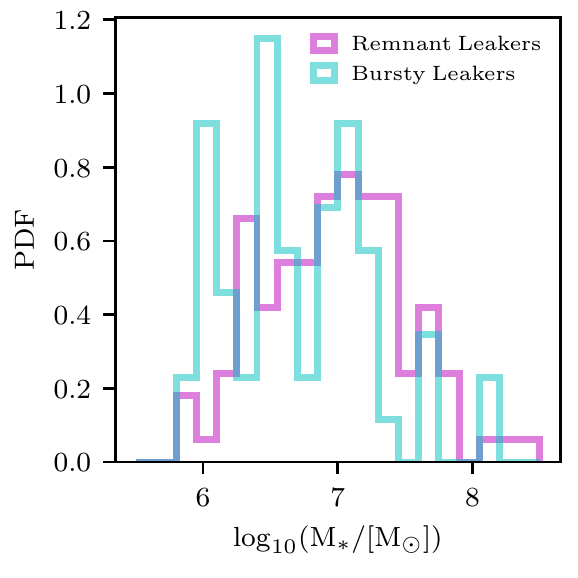}
\includegraphics[scale=1,trim={0 0.0cm 0cm 0cm},clip]{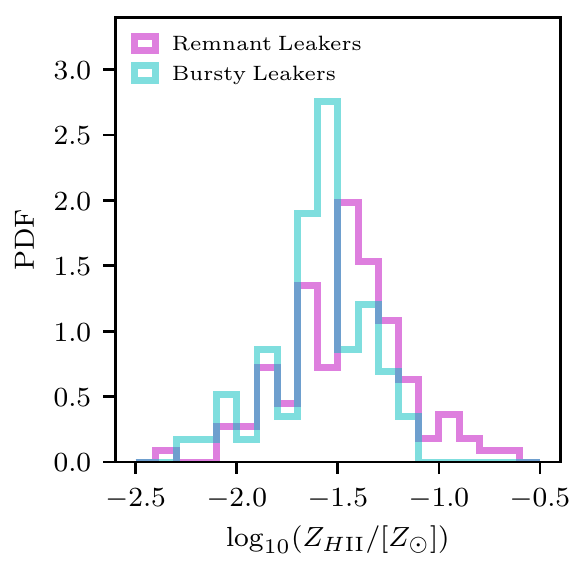}
}
\centerline{
\includegraphics[scale=1,trim={0 0.0cm 0cm 0cm},clip]{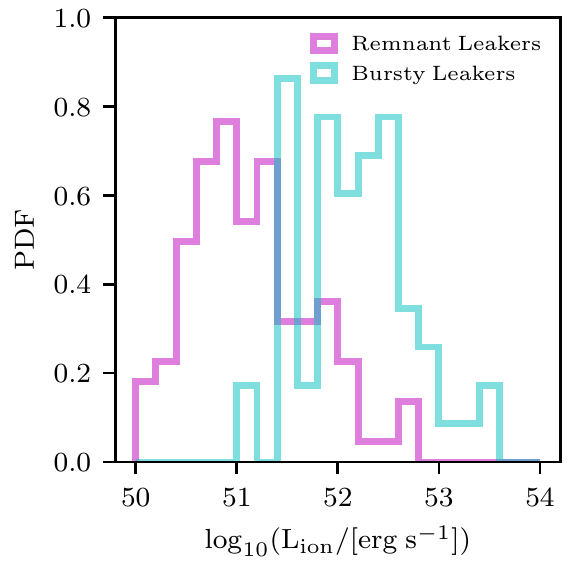}
\includegraphics[scale=1,trim={0 0.0cm 0cm 0cm},clip]{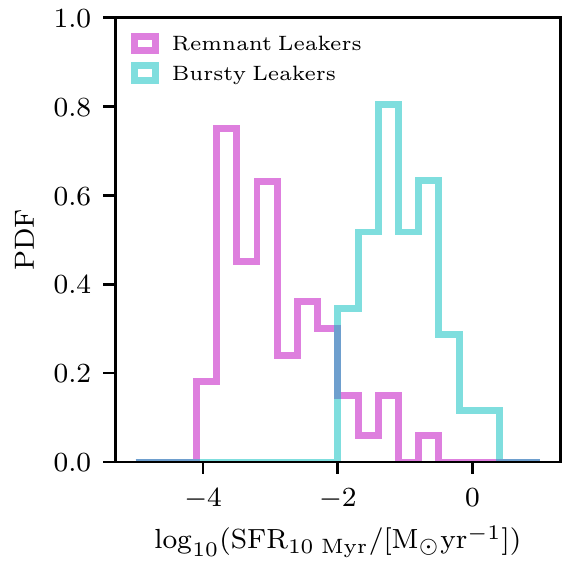}
\includegraphics[scale=1,trim={0 0.0cm 0cm 0cm},clip]{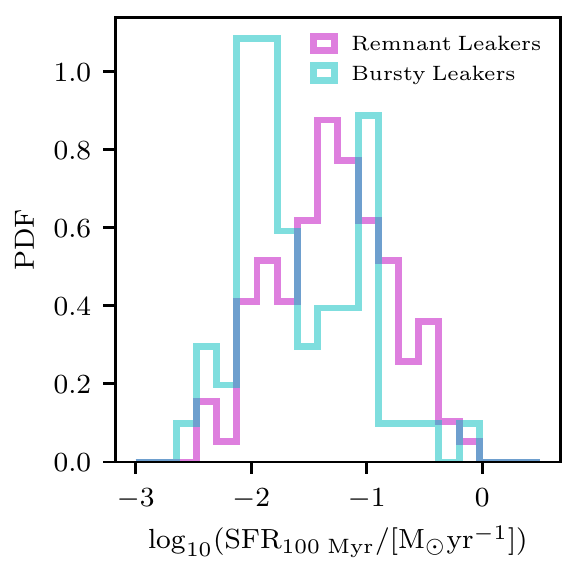}
}
\caption{Probability distribution functions (PDFs) comparing various galaxy and halo properties for the Bursty Leakers (cyan) with the Remnant Leakers (magenta) at $z=4.64$. We compare halo mass, stellar mass, H{\tiny II} region metallicity, total intrinsic ionizing luminosity, SFR$_{10}$, and SFR$_{100}$. Besides SFR$_{10}$ and ionizing luminosity, the two classifications of leakers have very similar properties because they represent similar galaxies in different evolutionary phases.}
\label{pop_hists}
\end{figure*}

\begin{figure*}
\centerline{
\includegraphics[scale=1,trim={0 0.0cm 0cm 0cm},clip]{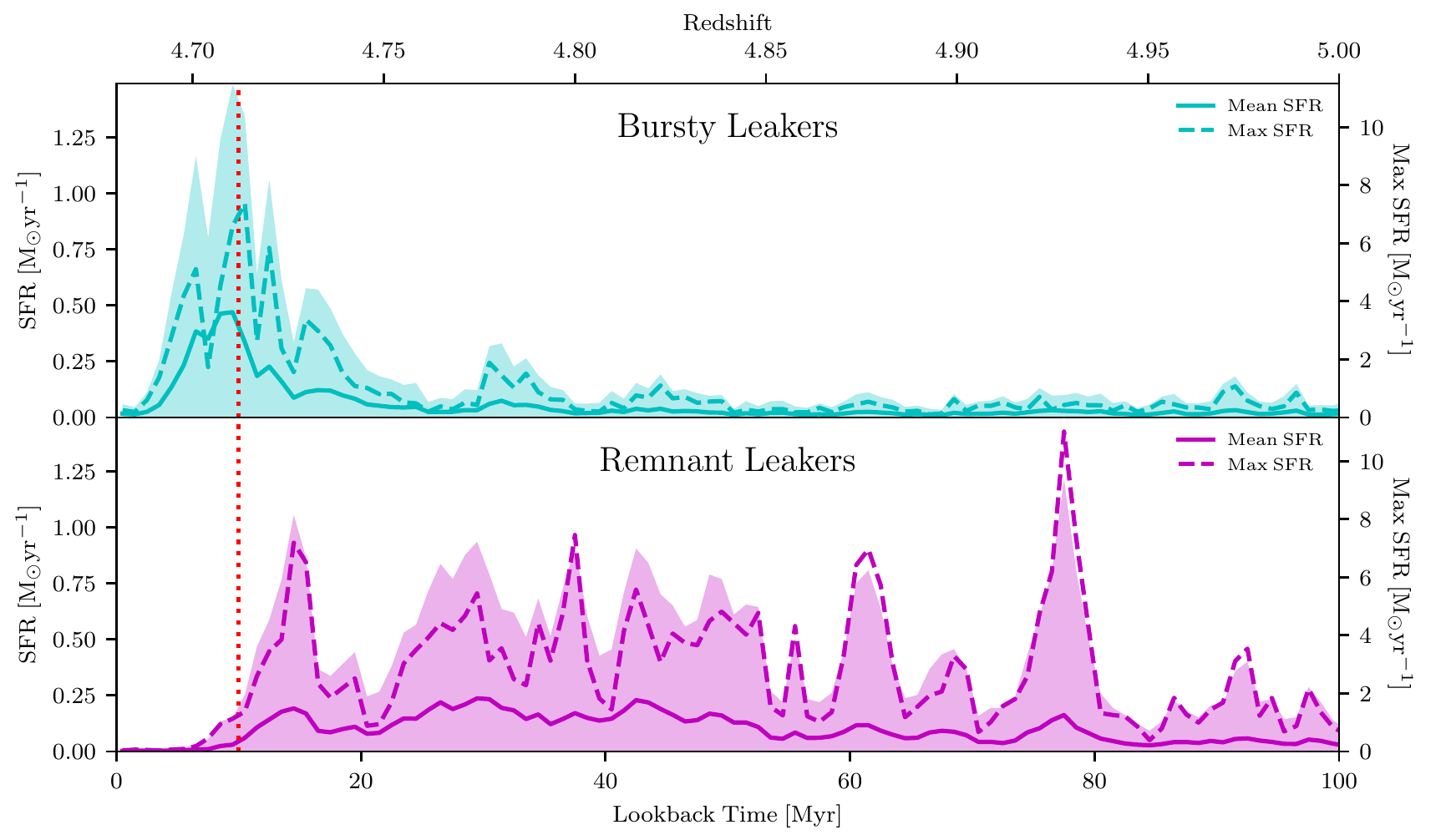}
}
\caption{Star formation histories of Bursty Leakers (top) and Remnant Leakers (bottom) for the 100~Myr prior to $z=4.64$. The solid line and shaded region show the mean and $1\sigma$ scatter of the SFR as a function of time with values as given on the y-axis on the left side of the plot. The dashed lines represent the maximum SFR among the population as a function of time and correspond to the y-axis on the right side of the plot. The vertical red dotted line denotes 10~Myr. We include all star particles within the virial radius of each halo and assume a $\Delta t$ of 1~Myr to compute the SFR. As populations, the Bursty Leakers exhibit very little star formation beyond 20~Myr since $z=4.64$, while the remnant leakers exhibit almost no star formation in the previous 10~Myr since $z=4.64$.}
\label{SFH_comp}
\end{figure*}

\section{Method}
\label{methods}
We employ the SPHINX$^{20}$ simulation \citep{Rosdahl2022}, the largest volume run of all simulations in the SPHINX suite of cosmological radiation hydrodynamics simulations \citep{Rosdahl2018,Katz2020,Katz2021,Katz2021b}. SPHINX$^{20}$ was run with the radiation hydrodynamics extension \citep{Rosdahl2013,Rosdahl2015} of the adaptive mesh refinement code {\small RAMSES} \citep{Teyssier2002}. The simulation incorporates state-of-the-art models for star formation \citep{Kimm2017} and stellar feedback \citep{Kimm2015} and employs the variable speed of light approximation \citep{Katz2017} to capture the motion of ionization fronts through the ISM and IGM. With a maximum spatial resolution of $\sim10\ {\rm pc}$, the simulation is able to model a multi-phase ISM structure and the low density channels through which LyC photons escape. The large volume of 20$^3$~cMpc$^3$ allows us to sample a wide range of galaxy properties, resolving haloes of $10^8\ {\rm M_{\odot}}$ by 400 dark matter particles. Full details of the physics included in the simulations is described in detail in \cite{Rosdahl2022} and our methods for calculating line emission, in particular [C{\small II}], as well as any minor changes between SPHINX$^{10}$ and SPHINX$^{20}$ are described in \cite{Katz2021b}. Escape fractions are calculated in post-processing by using Monte Carlo radiative transfer \citep[{\small RASCAS},][]{Rascas2020} to follow 912\angstrom~photons from star particles to the virial radius of each halo (see \citealt{Katz2021b}). Not all simulations use the same radius to measure the escape fraction; however, our choice is consistent with all other work on the {\small SPHINX} simulations \citep{Rosdahl2022}. For each galaxy $10^7$ photon packets are distributed among the star particles with initial positions randomly sampled from a multinomial distribution based on the location and ionizing emissivity. We measure escape fractions along individual lines of sight as well as the angle-averaged values. By measuring $f_{\rm esc}$ in two ways, we can better constrain the impact of anisotropic leakage in prospective observations. However, we note that the angle-averaged value is the important quantity for measuring the impact of individual galaxies on reionization.

In this work, we primarily study the $z=4.64$ snapshot, the final snapshot of the simulation, for which we calculated [C{\small II}] emission for the nearly 30,000 galaxies with halo masses $\geq10^8\ {\rm M_{\odot}}$. There are significantly more galaxies observed in [C{\small II}] at $z<6$ and, due to the neutral IGM at high redshift, direct LyC detections are only possible at $z<6$. This motivates our study of the lowest redshift snapshot of the simulation. However, for our results to matter for reionization, we must also show that the trends between $f_{\rm esc}$ and galaxy properties at $z=4.64$ also hold for $z\geq6$. For this reason, we will then link the results to $z=6$ where the simulation resolves similar numbers of galaxies. 

Emission lines are calculated on a cell-by-cell basis in post-processing by running {\small CLOUDY} models \citep{Ferland2017} on all cells in the simulation based on their gas density, metallicity, temperature, dust content, and local radiation field. We have adopted the solar abundance pattern model from \cite{Katz2021b}. Due to the large volume and high resolution of the simulation, at $z=4.64$, halo masses range from $10^8~{\rm M_{\odot}}-10^{11.7}~{\rm M_{\odot}}$, stellar masses span $10^{2.5}~{\rm M_{\odot}}-10^{10.5}~{\rm M_{\odot}}$, SFRs (averaged over 10~Myr) vary between $0~{\rm M_{\odot}yr^{-1}}-10^{1.9}~{\rm M_{\odot}yr^{-1}}$, and finally H{\small II} region metallicities range between $10^{-3.3}~Z_{\odot}-10^{0.1}~Z_{\odot}$. We expect these values to be rather typical of any 20$^3$~cMpc$^3$ volume as the initial conditions in the simulation were chosen so that the $z=6$ halo mass function was most representative of the average of multiple random realizations.

\begin{figure}
\centerline{
\includegraphics[scale=1,trim={0 0.0cm 0cm 0cm},clip]{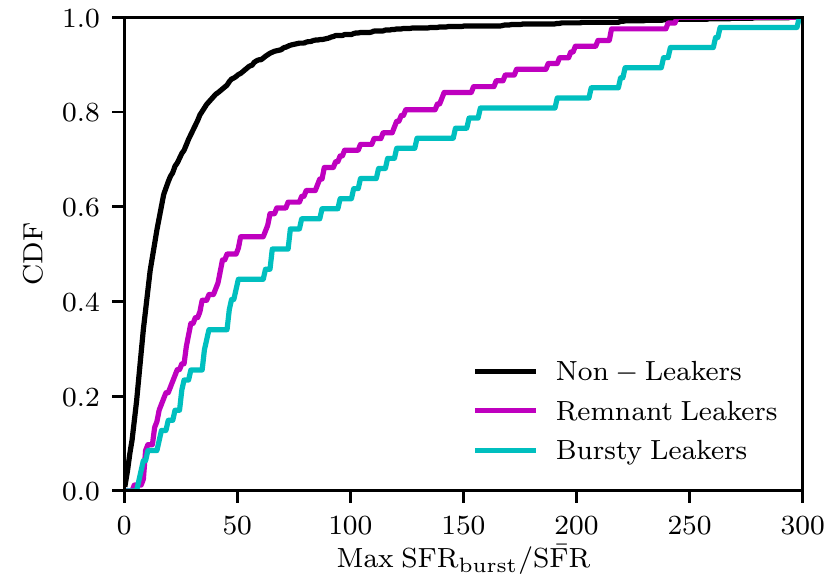}
}
\caption{Cumulative distribution function of the maximum SFR in the burst to the typical SFR of the galaxy prior to the burst for Bursty leakers (cyan), Remnant leakers (magenta), and non-leakers (black.)}
\label{sfr_cdf}
\end{figure}

\begin{figure}
\centerline{
\includegraphics[scale=1,trim={0 0.0cm 0cm 0cm},clip]{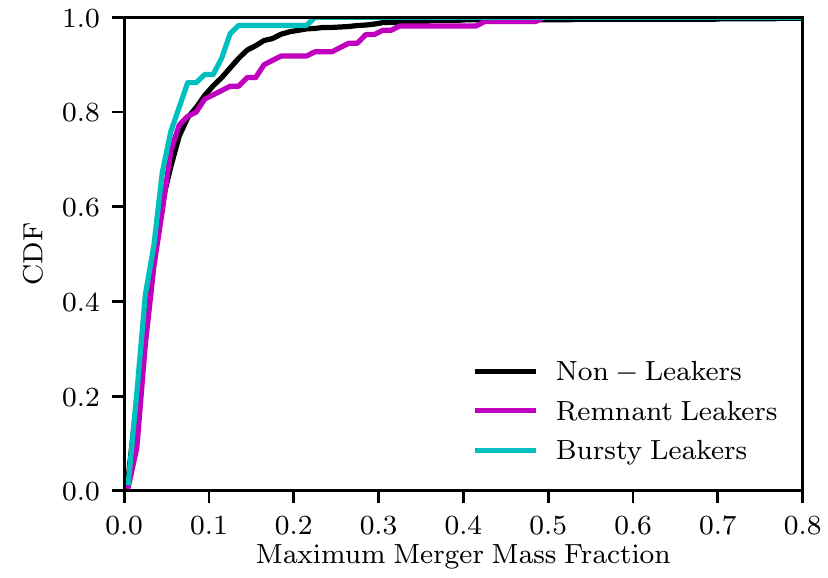}
}
\caption{Cumulative distribution function of the maximum fractional change in dark matter mass in the 100~Myr time period before $z=4.64$ for Bursty leakers (cyan), Remnant leakers (magenta), and non-leakers (black.)}
\label{merger}
\end{figure}

\begin{figure}
\centerline{\includegraphics[scale=1,trim={0 0.0cm 0cm 0cm},clip]{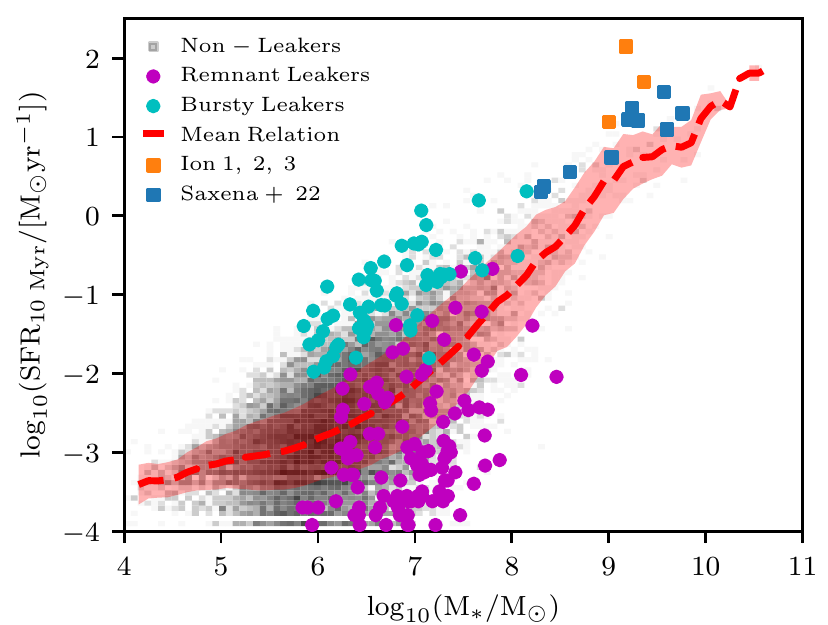}}
\centerline{\includegraphics[scale=1,trim={0 0.0cm 0cm 0cm},clip]{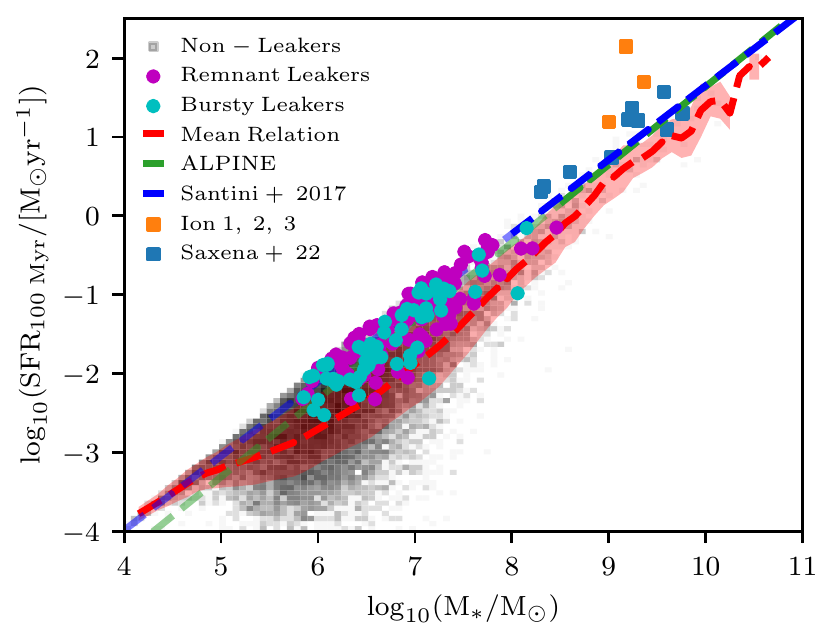}}
\caption{Star formation main sequence using SFR$_{10}$ (top) or SFR$_{100}$ (bottom) for SPHINX$^{20}$ galaxies at $z=4.64$. Non-Leakers are shown as the grey histogram while Bursty and Remnant Leakers are shown in cyan and magenta, respectively. The dashed red line and shaded region represent the mean and $1\sigma$ scatter for SPHINX$^{20}$ galaxies. In the bottom panel, we compare our simulated main sequence with observational constraints from \protect\cite{Santini2017} and \protect\cite{Khusanova2021} (ALPINE) and find good agreement, especially at high stellar masses. The lines become translucent at low stellar masses where observational inferences have been extrapolated. We also show known $z=3-3.5$ LyC leakers from \protect\cite{Saxena2022} and {\it Ion 1, 2} and {\it 3} from \protect\cite{Vanzella2012,Vanzella2016,Vanzella2018} to demonstrate that all fall above the main sequence. Note that there are no direct constraints on SFR$_{10}$ and SFR$_{100}$ for the observed samples as they are derived from the SED and UV, so we show them only for qualitative comparison.}
\label{smass_sfr}
\end{figure}

We classify SPHINX$^{20}$ galaxies into four different groups based on their angle-averaged LyC escape fractions and star formation histories: Non-Leakers, Intermediate Leakers, Bursty Leakers, and Remnant Leakers. Non-leakers are defined as galaxies with angle-averaged $f_{\rm esc}\leq5\%$, Intermediate Leakers have $5\%<f_{\rm esc}<20\%$, Bursty Leakers have $f_{\rm esc}\geq20\%$ and their star formation rate (SFR) averaged over the previous 10~Myr (SFR$_{10}$) is greater than their SFR averaged over the previous 100~Myr (SFR$_{100}$), and finally, Remnant Leakers have $f_{\rm esc}\geq20\%$ and ${\rm SFR}_{100}\geq{\rm SFR}_{10}$\footnote{In practice, there is a smooth transition between the Bursty and Remnant Leaker populations in terms of SFR$_{10}$/SFR$_{100}$, but for simplicity, we have introduced a hard cutoff.}. A schematic of this distribution can be seen in Figure~\ref{classify}. It should be noted that neither SFR$_{10}$ nor SFR$_{100}$ are observable quantities. They must be inferred from observations, for example by fitting the SED or comparing with IR or line emission, with each indicator potentially probing star formation on different time scales \citep[e.g.][]{Calzetti2007,Kennicutt2012}.  We will show that the trends we find for Bursty and Remnant leakers should persist if we replace SFR$_{10}$ and SFR$_{100}$ with SFR indicators that are sensitive to star formation on very short and slightly longer time scales. H$\alpha$ luminosity and infrared luminosity are such indicators \citep[e.g.][]{Kennicutt1992}.

Our classification scheme is related to the different modes of $f_{\rm esc}$. Bursty Leakers are subject to strong photoionization feedback as well as SNe feedback from the most massive stars. The Remnant Leakers have had a burst of star formation sometime in the recent past compared to their current SFR. It is likely that the SNe feedback from the previous burst disrupted the ISM enough to shut down star formation. Interestingly, the galaxies with SFR$_{10}$/SFR$_{100}<2\times10^{-3}$ are all Remnant Leakers with $f_{\rm esc}>20\%$. Thus the first mode is high $f_{\rm esc}$ due to radiation and early SN feedback while the latter mode is after significant SN feedback has occurred. Note that our two modes are distinct from whether the radiation leaks through holes or more uniformly due to $\tau<1$ \citep[e.g.][]{Zackrisson2013}; however, there are parallels that we discuss below.

The cut at an escape fraction of 20\% is arbitrary and about twenty times\footnote{The luminosity-weighted escape fraction in SPHINX$^{20}$ at $z=4.64$ is 1\%. This is consistent with the upper limits on the ``average'' escape fraction at $z=3.3$ measured by \citet{Grazian2017} using ultra-deep U-band imaging.} as much as the global (i.e. ionizing luminosity-weighted) escape fraction at $z=4.64$ in SPHINX$^{20}$; however, our results are not fundamentally different if we employ other thresholds\footnote{We have tested that all of the trends we present hold down to an $f_{\rm esc}$ threshold of 5\% (i.e. removing the intermediate leaker bin completely). Similar thresholds were recently employed in observational studies of Ly$\alpha$ emitters at low-redshift \citep[e.g.][]{Naidu2022,Matthee2022}. Using a threshold $\gg20\%$ results in too few galaxies being classified as leakers. The number of leakers in each class decreases approximately linearly when varying the threshold $f_{\rm esc}$ from 5\% to 20\%.}. In total, at $z=4.64$, selecting only galaxies with [C{\small II}] luminosities $>10^2L_{\odot}$ and ${\rm SFR_{10}}>10^{-4}\ {\rm M_{\odot}yr^{-1}}$ we find 58 Bursty Leakers, 111 Remnant Leakers, 695 Intermediate Leakers, and 16,085 Non-Leakers.

\section{Results}
\label{results}

\subsection{Galaxy Property Comparison of Bursty and Remnant Leakers}
In Figure~\ref{pop_hists} we compare various galaxy properties of the two classes of leakers. Bursty and Remnant leakers have very similar distributions of halo mass, stellar mass, and H{\small II} region\footnote{Throughout this paper, H{\tiny II} regions are defined as gas cells with H{\tiny II} fractions $>50\%$, temperatures $<10^{5.5}\ {\rm K}$, and gas densities $>1\ {\rm cm^{-3}}$.} metallicity. Most of the leakers in the simulation have halo masses of $\sim10^9\ {\rm M_{\odot}}$. More massive galaxies in SPHINX$^{20}$ are less likely to be leakers. The 10~Myr-averaged SFRs of the Bursty Leakers are significantly greater than the Remnant Leakers, which is unsurprising given how the two populations are defined. More specifically, the median SFR$_{10}$ of the Bursty Leakers is $0.056\ {\rm M_{\odot}yr^{-1}}$, more than an order of magnitude greater than the median of $0.001\ {\rm M_{\odot}yr^{-1}}$ for the Remnant Leakers. In contrast, both Bursty and Remnant Leakers exhibit very similar\footnote{The median SFR$_{100}$ for the Bursty leakers is slightly smaller (i.e. 70\%) of the median value of the Remnant Leakers. This is well within the sampling uncertainty of the two distributions as the tension is $<1\sigma$.} 100~Myr-averaged SFRs, indicating that the total amount of star formation over the past 100~Myr is also similar for the two populations. Because of the significantly enhanced recent star formation, Bursty Leakers also exhibit much higher ionizing luminosities compared to remnant leakers. We find more than an order of magnitude difference between the median ionizing luminosities of the two groups of galaxies. 

\begin{figure*}
\centerline{\includegraphics[scale=1,trim={0 0.0cm 0cm 0cm},clip]{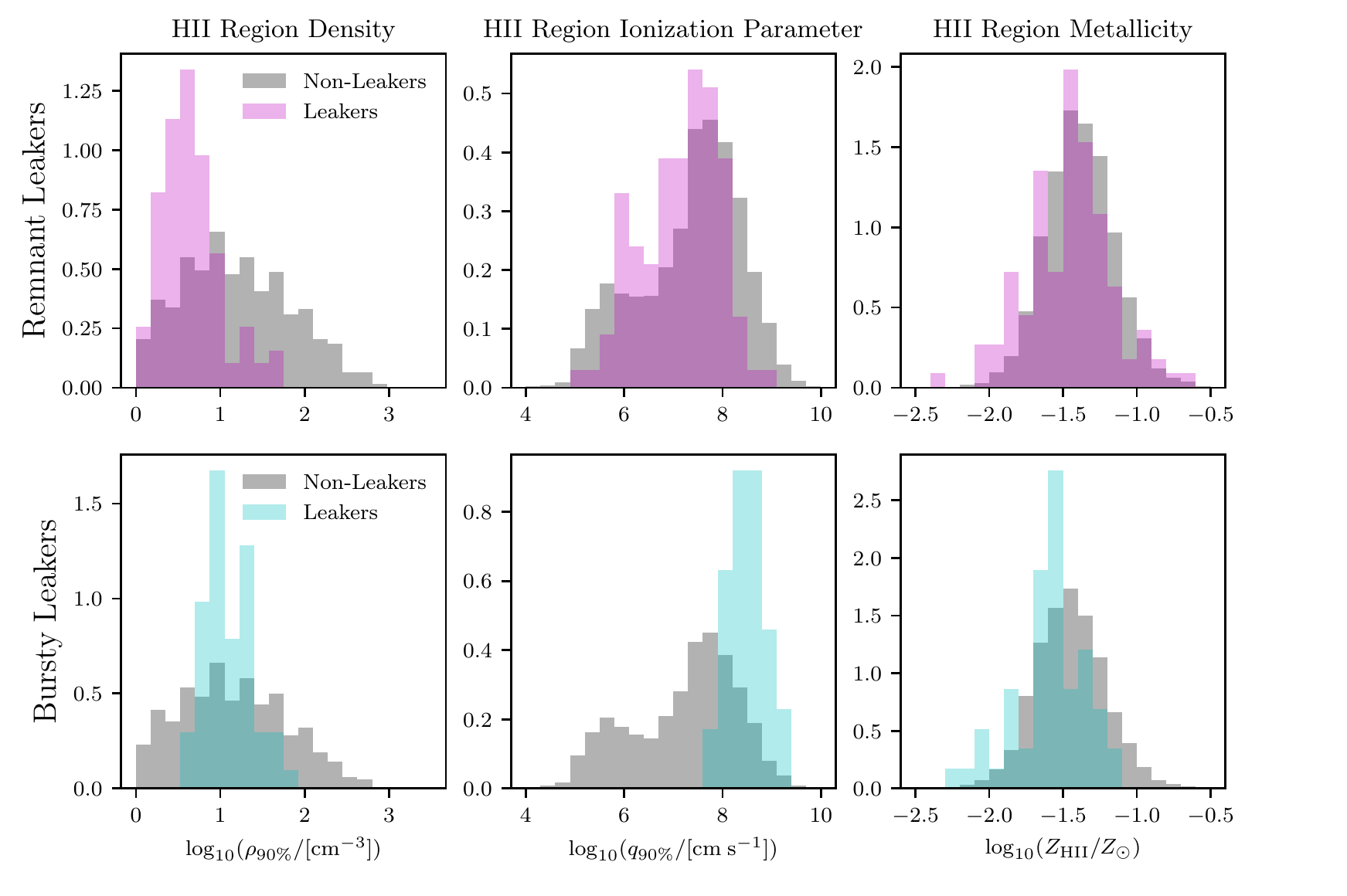}}
\caption{Comparison of ISM properties of Remnant Leakers (top, magenta) and Bursty Leakers (bottom, cyan) with control samples of Non-Leakers that were stellar mass-selected to exhibit the same distribution of stellar masses as each leaker population. From left to right, we compare probability distribution functions of 90th percentile H{\tiny II} region gas density ($\rho_{90\%}$), 90th percentile H{\tiny II} region ionization parameter ($q_{90\%}$, defined as ionizing photon flux divided by gas density), and H{\tiny II} region metallicity.}
\label{HII_region}
\end{figure*}

In summary, Bursty Leakers and Remnant Leakers exhibit very similar fundamental galaxy properties except for the fact that the burst of star formation responsible for the LyC leakage is currently underway or happened very recently for the Bursty Leakers, whereas it happened further in the past for Remnant Leakers. 

We continue the comparison in Figure~\ref{SFH_comp} where we show the mean and maximum SFRs as a function of time in the 100~Myr period prior to $z=4.64$ for the Bursty Leakers (top) and the Remnant Leakers (bottom). In both populations of leakers, star formation has significantly decreased at $z=4.64$, signifying the impact of strong stellar feedback (both radiative and SNe). The peak in the mean SFR for the Bursty Leakers occurred $\sim8$~Myr prior to $z=4.64$; however for individual leakers in this population, the maximum SFR could have occurred up to 18~Myr prior to the snapshot. In contrast, the mean SFR of the Remnant Leakers remains relatively constant between 15~Myr and 80~Myr prior to $z=4.64$ due to the fact that we are averaging the star formation histories of multiple galaxies that have had bursts at different times.

In both populations, the galaxies undergo extreme bursts of star formation which is highly correlated with having a high escape fraction. The primary difference between the two is the timing of the burst. The severity of the burst can be calculated by separating the SFR in the range $\pm10$~Myr before and after the maximum recorded SFR and comparing the SFR in the burst to the typical SFR of the galaxy in the 50~Myr time period prior\footnote{The choice of time window to measure the ``typical'' SFR of a galaxy is arbitrary. We have experimented with choosing windows that account for the periods both before an after the burst and our results are qualitatively similar. We have chosen to only measure the SFR prior to the burst in order to remove the time period where star formation is significantly suppressed due to the burst.} to the burst. In this experiment we measure the SFRs over 1~Myr intervals. For both Bursty and Remnant Leakers, the maximum SFRs in the burst are $\sim90\times$-$100\times$ the typical SFR of the galaxy (as given by the median of the two populations). Rather than taking the maximum SFR in the burst, we can compare the mean SFR in the burst window and still find that the enhancement in SFR is $\sim14\times$ the typical SFR of the galaxy.

In order to estimate the probability of having such a strong burst of star formation, we construct a control sample of non-leakers by randomly selecting $\sim1000$ galaxies such that the stellar mass distribution of this control sample is the same as the combination of Bursty and Remnant Leakers. The maximum SFRs in the burst of control sample galaxies are typically only $13\times$ the typical SFR and the median SFR in the burst window is only $3\times$ the typical SFR. This can easily be seen in Figure~\ref{sfr_cdf} where we show the cumulative distribution function of the ratio of the maximum SFR in the burst to the typical SFR of the galaxy prior to the burst for Bursty leakers, Remnant leakers, and non-leakers. Here we see that the strength of the bursts for the Bursty and Remnant leakers are, in general, significantly stronger than what is observed for non-leakers. This confirms previous claims \citep[e.g.][]{Trebitsch2017,Kimm2017} that $f_{\rm esc}$ is feedback regulated and extreme bursts of star formation are required to disrupt the ISM enough so that LyC photons can efficiently escape. Among the control sample, only 6\% of galaxies exhibit maximum SFRs as bursty as what is found for the leaker sample in the 100~Myr prior to $z=4.64$. Similarly only 8\% of galaxies in the control sample exhibit mean SFRs in the 20~Myr burst window that are $\sim13\times$ the typical SFR of the galaxy. Both statistics indicate star bursts of this strength are rare. 

Because mergers are more common at high redshift \citep{Hopkins2010}, it is interesting to consider whether mergers are responsible for the starbursts that lead to high $f_{\rm esc}$. To assess this, we look in the 100~Myr window prior to $z=4.64$ and record the maximum fractional change in dark matter mass (i.e. $(M_{{\rm DM},t+1}-M_{{\rm DM},t})/M_{{\rm DM},t}$) between simulation outputs. Galaxies with large fractional changes close to 1 will have undergone a near equal mass merger. In Figure~\ref{merger} we show the cumulative distribution function of these maximum fractional changes for the Bursty leakers, Remnant leakers, and the randomly selected 1000 non-leakers. We do not find a significant excess of mergers in the Bursty or Remnant leakers compared to the non-leakers. For all three galaxy populations, the maximum merger mass ratio was 20:1 or less for $\sim50\%$ indicating that a major merger is not required for high $f_{\rm esc}$.

The fact that the Non-Leaker population contains some star bursts as strong as what we see in the leaker population introduces the question of why some star bursts lead to high $f_{\rm esc}$ while others do not. 30\% of the sample of Non-Leakers that exhibit strong starbursts have either had a very recent burst (within the past 5~Myr) or the star burst occurred in the range between $70-100$~Myr prior to $z=4.64$. In the former case, feedback has not had enough time to clear channels in the H{\small I} distribution. In the latter, there may have been enough time for the galaxy to re-collapse and form a dense ISM structure with a low escape fraction. For the remaining 70\%, there are two possibilities. Either the haloes never exhibited a high $f_{\rm esc}$, despite the star burst, or the star burst did efficiently clear channels in the ISM but the gas re-collapsed and settled much faster than what we see in the Remnant Leaker population, thus their $f_{\rm esc}$ just happens to be low at $z=4.64$. As we are primarily interested in developing methods to find current LyC leakers and constrain the mechanisms by which photons escape, we leave this question open for future work.

Another way to quantify these bursts is by comparing the Bursty and Remnant Leakers with the Non-Leaker galaxy population on the galaxy formation main sequence (i.e. stellar mass versus SFR). In the top panel of Figure~\ref{smass_sfr} we show stellar masses of SPHINX$^{20}$ galaxies at $z=4.64$ compared with SFR$_{10}$. In general, the Bursty Leakers populate a region significantly higher than the main sequence while the Remnant Leakers mostly fall significantly below. As we discussed earlier, the Bursty Leakers have had large bursts of star formation in the past 10~Myr while the Remnant Leakers have had their star formation nearly completely shut down so it is unsurprising that we find strong differentiation on the galaxy main sequence. In contrast, if we show the main sequence but replace SFR$_{10}$ with SFR$_{100}$, as shown in the bottom panel of Figure~\ref{smass_sfr}, both populations of leakers now fall above the mean relation. This is because both Bursty and Remnant Leakers on average have exhibited extreme bursts of star formation in the past 100~Myr. Consistent with our simulations, candidate LyC leakers at $z=3-3.5$ with \fesc $>0.2$ from \cite{Saxena2022} as well as other strong LyC leaking galaxies, \emph{Ion 1, 2} and \emph{3} \citep{Vanzella2012, Vanzella2016, Vanzella2018}, also fall above the main sequence. 

We have compared our simulated galaxy main sequence using SFR$_{100}$ with observational estimates from the ALPINE survey \citep{Khusanova2021}. Data in the IR was used to make estimates of SFR, which is comparable to star formation over the last 100~Myrs. We find very good agreement between our simulations and the observational constraints. The key result from this exercise is that the location of a LyC leaker on the galaxy formation main sequence is highly dependent on the time scale over which an SFR indicator is sensitive. 

\begin{figure*}
{\huge Bursty Leakers}
\centerline{
\includegraphics[scale=1,trim={0 0.0cm 0cm 0cm},clip]{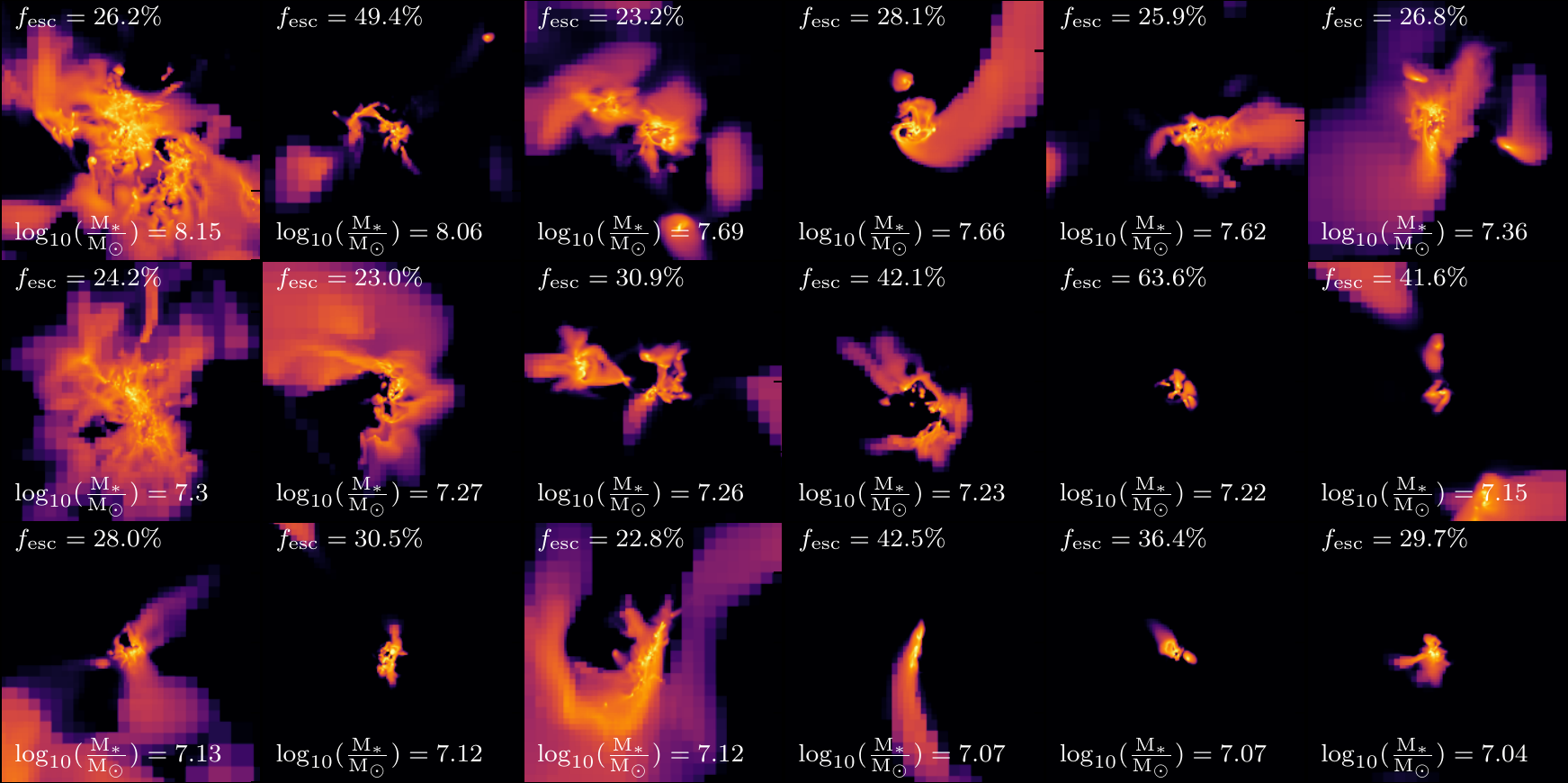}
}
{\huge Remnant Leakers}
\centerline{
\includegraphics[scale=1,trim={0 0.0cm 0cm 0cm},clip]{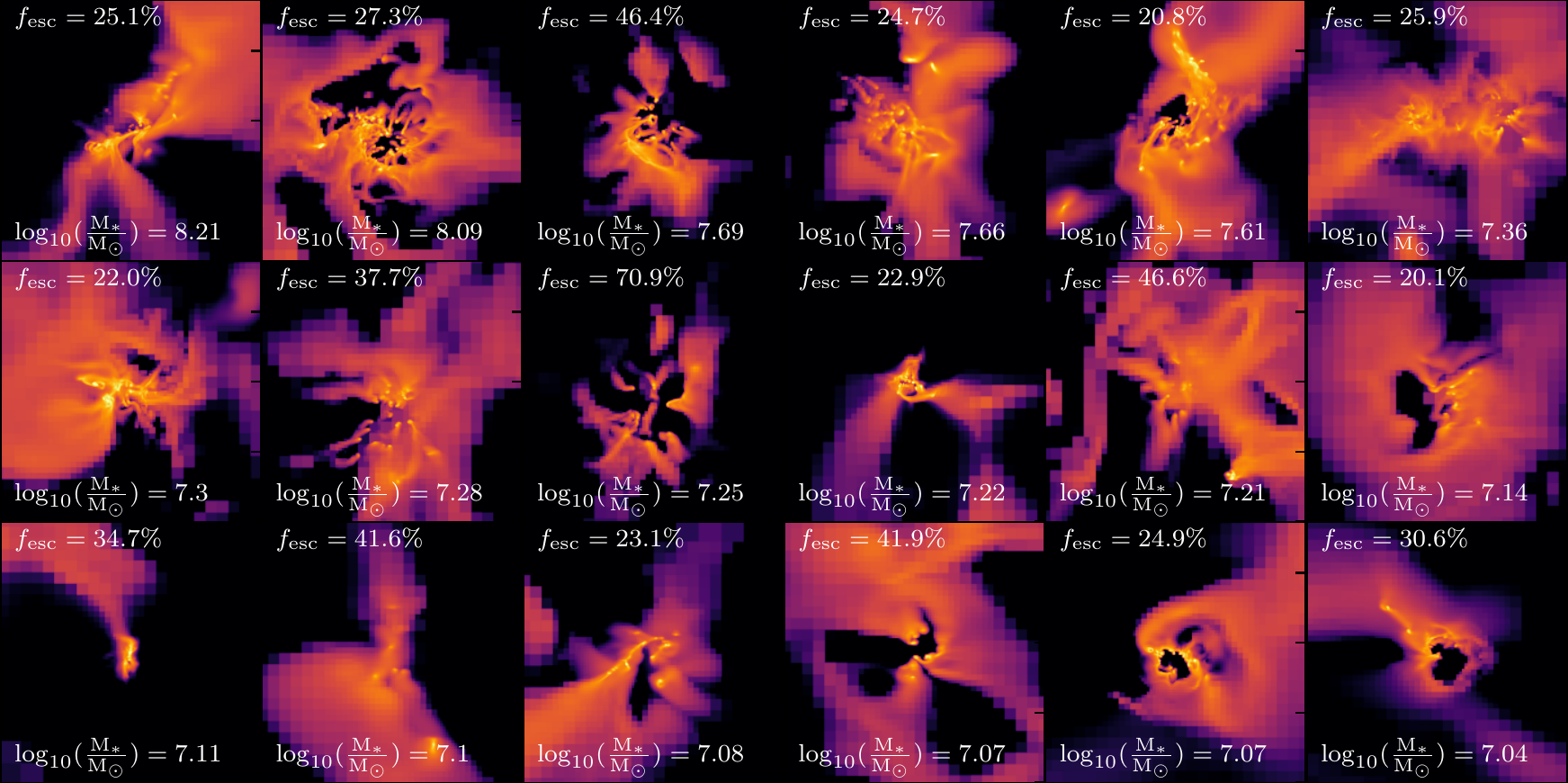}}
\centerline{\includegraphics[scale=1,trim={0 0.0cm 0cm 0cm},clip]{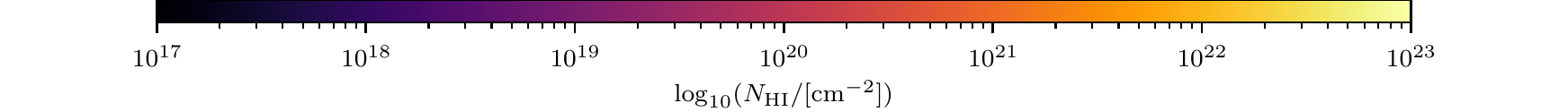}}
\caption{HI column density maps for Bursty Leakers (top) and Remnant Leakers (bottom) at $z=4.64$. All maps are 10~physical kpc in width and we list the escape fraction and stellar mass of each object. We show the densest Bursty Leakers (representing 31\% of the total sample) with a corresponding stellar mass selected sample of Remnant Leakers. In the top panel, we see that the ISM is still intact for many Bursty Leakers but the strong radiation field has often ionized a significant amount of the gas in the CGM. In contrast, Remnant Leakers tend to show large holes is a disrupted ISM.}
\label{leaker_maps}
\end{figure*}

\begin{figure*}
\centerline{
\includegraphics[scale=1,trim={0 0.0cm 0cm 0cm},clip]{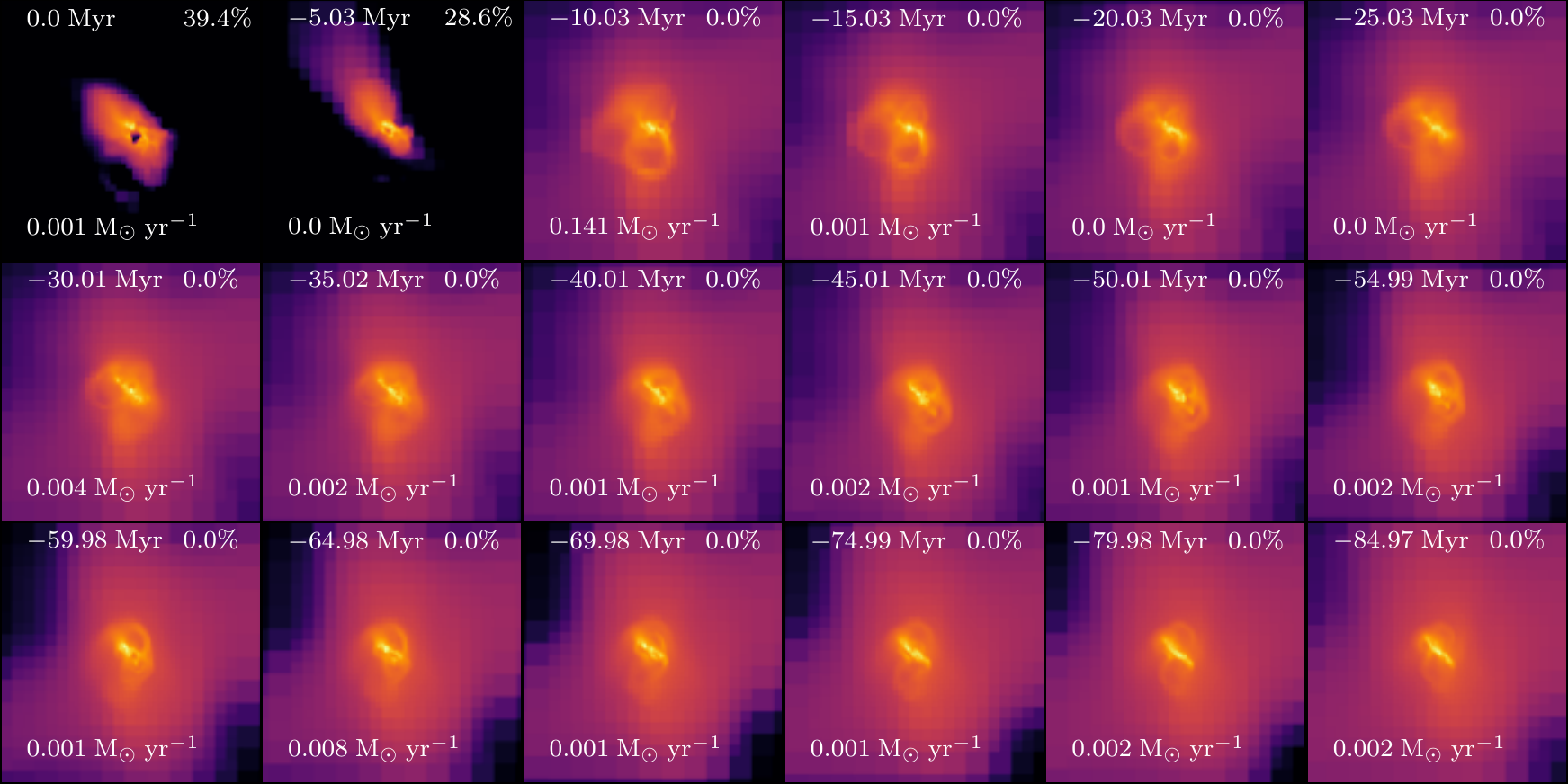}
}
\centerline{
\includegraphics[scale=1,trim={0 0.0cm 0cm 0cm},clip]{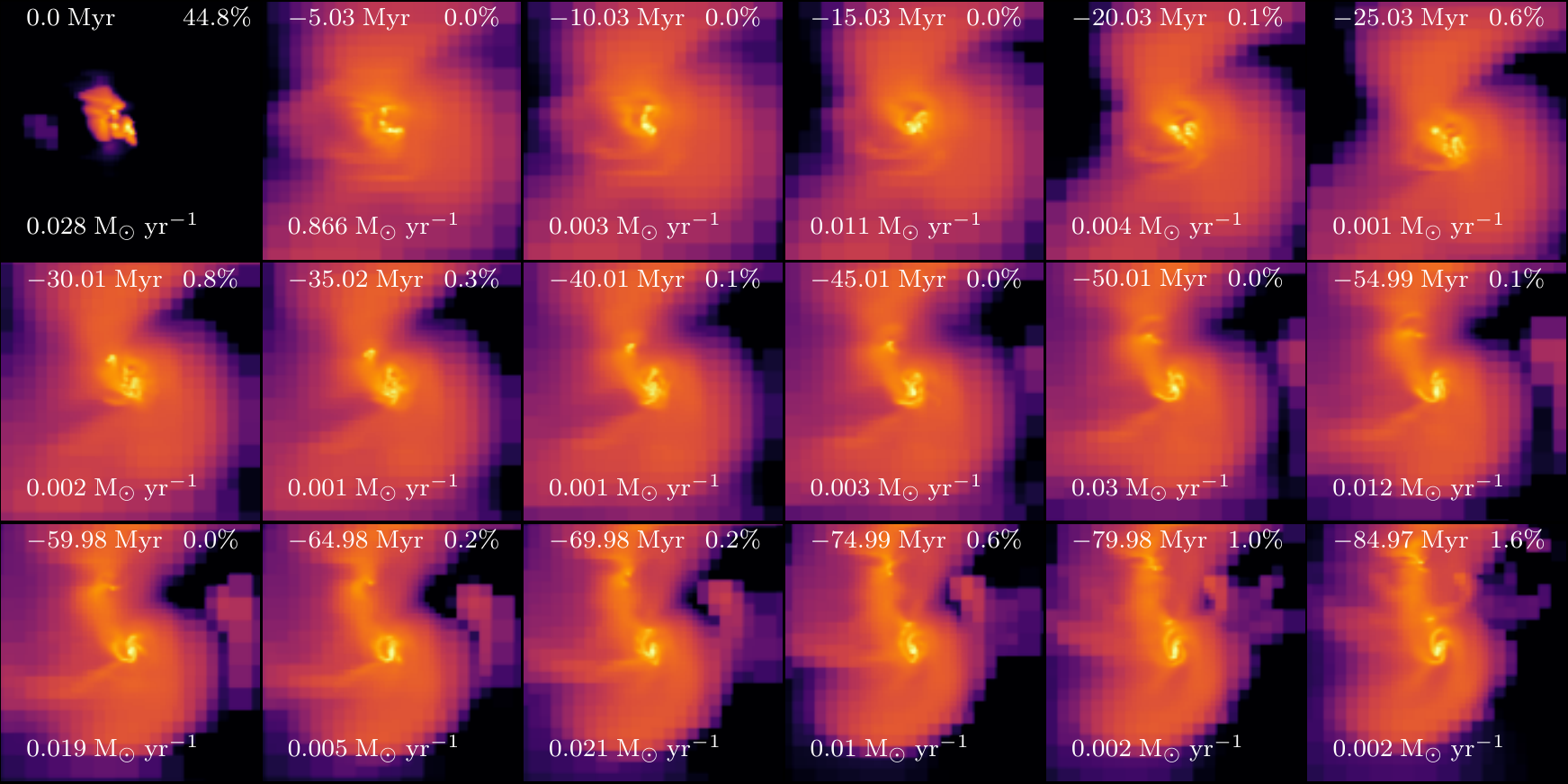}}
\centerline{\includegraphics[scale=1,trim={0 0.0cm 0cm 0cm},clip]{./figures/cbar_leaker_maps.pdf}}
\caption{Maps for the HI column density in a time series of the 85~Myr prior to $z=4.64$ for two Bursty Leakers with halo masses of $10^{8.3}\ {\rm M_{\odot}}$ and $10^{9.1}\ {\rm M_{\odot}}$ and stellar masses of $10^{6.1}\ {\rm M_{\odot}}$ and $10^{6.7}\ {\rm M_{\odot}}$, respectively. The three labels indicate the time of the snapshot, the LyC escape fraction, and the star formation rate measured over the previous 1~Myr.}
\label{leaker_time}
\end{figure*}

\begin{figure*}
\centerline{
\includegraphics[scale=1,trim={0 0.0cm 0cm 0cm},clip]{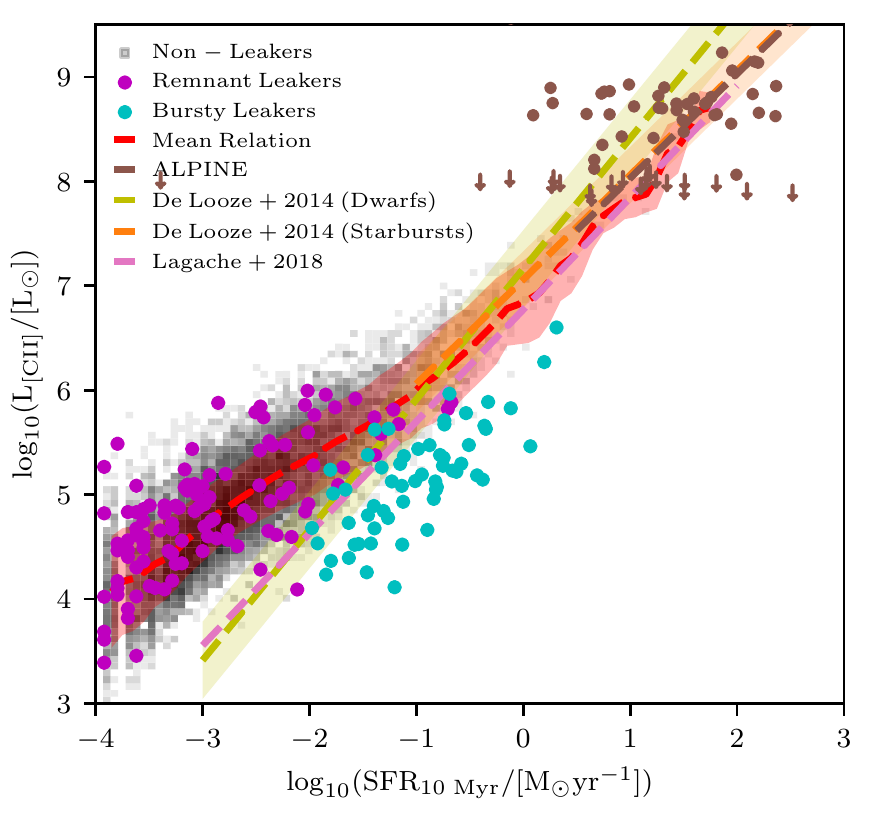}
\includegraphics[scale=1,trim={0 0.0cm 0cm 0cm},clip]{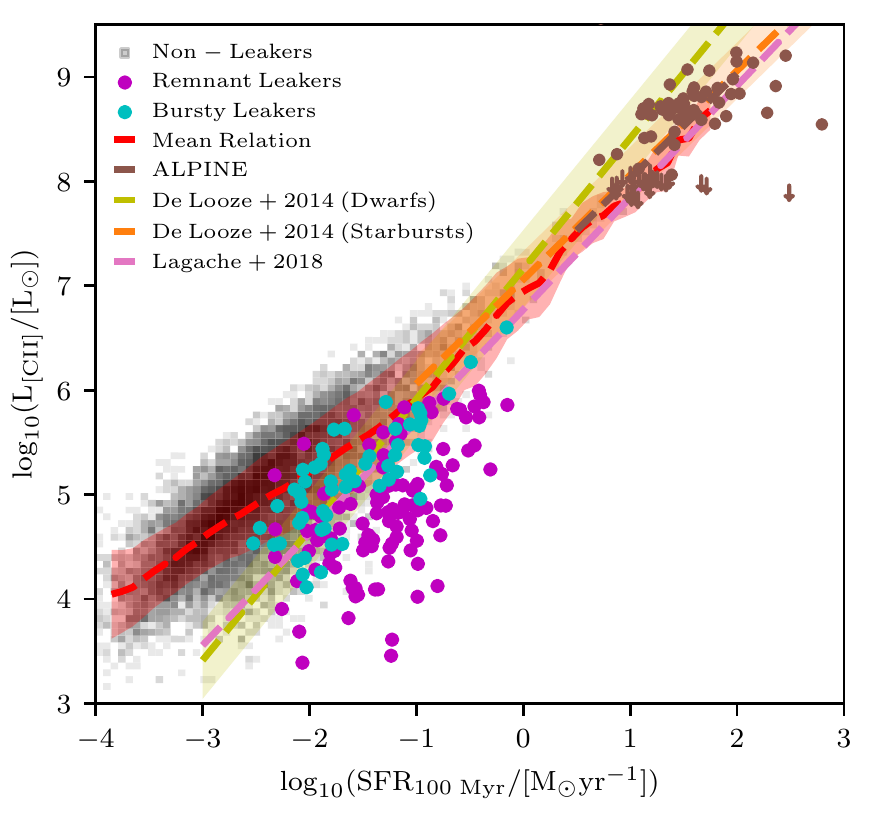}
}
\caption{[C{\small II}]-SFR$_{10}$ (left) and [C{\small II}]-SFR$_{100}$ (right) relation for {\small SPHINX$^{20}$} galaxies at $z=4.64$ compared with fits to various types of low-redshift galaxies \protect\citep[olive-green and orange,][]{DeLooze2014} as well as high-redshift observations \protect\citep[brown,][]{Schaerer2020} and models \protect\citep[pink,][]{Lagache2018}. Bursty Leakers are shown in cyan while Remnant Leakers are shown in magenta. All SPHINX$^{20}$ Non-Leakers are shown as the grey background histogram. The mean relation from the simulation is shown as the red dashed line and the red shaded region represents the $1\sigma$ scatter. Individual $z<5.5$ galaxies from ALPINE are shown in brown with points representing detections and arrows representing $3\sigma$ upper limits on [C{\small II}] luminosity. ALPINE SFRs that are calculated from H$\alpha$ inferred from \emph{Spitzer} photometry and using SED fitting \citep{Faisst2020} against their corresponding [C{\small II}] luminosities \citep{Bethermin2020} are shown in the left and right panels, respectively.}
\label{line_sfr}
\end{figure*}

\subsection{ISM Property Comparison of Bursty and Remnant Leakers}
The mechanism (i.e. radiation-bounded with holes or density bounded) by which photons are leaking in each population of LyC leaker is still unclear. One method for elucidating this physics is to compare the ISM properties of each population conjointly as well as with control samples of Non-Leakers that are selected to exhibit the same stellar mass distribution as each leaker population\footnote{We note that there is no clear method for how to properly choose a control sample. We have controlled for stellar mass but it might also be appropriate to control for star formation history, star formation rate, halo mass, or any other galaxy property. }. 

In Figure~\ref{HII_region} we compare distributions of 90th percentile\footnote{Describing the properties of the ISM by a single number is not a well defined problem. We have chosen the 90\% percentile value of the distribution for the gas density and ionization parameter to be consistent with \protect\cite{Katz2021b}. The results are not fundamentally different if other percentiles are used. The values are computed by locating all gas cells that are part of the H{\small II} regions of the galaxies and measuring the 90th percentile values of the distribution.} H{\small II} region gas density ($\rho_{90\%}$), 90th percentile H{\small II} region ionization parameter ($q_{90\%}$), and H{\small II} region metallicity with the control samples of stellar mass-selected Non-Leaker galaxies. Beginning with similarities, we find no difference in H{\small II} region metallicity between any of the samples. The peaks of the distribution occur at $\sim3\%~Z_{\odot}$. Where our stellar masses overlap, the mass-metallicity relation of SPHINX$^{20}$ galaxies is in good agreement with observational estimates from \cite{Faisst2016} at $z=5$ so we expect our predicted high-redshift leaker metallicities to be reasonably robust. We find no differentiation between the Bursty Leakers, Remnant Leakers, and Non-Leakers on the stellar mass-metallicity relation (not shown).

Continuing with the differences, the middle panels of Figure~\ref{HII_region} compare the ionization parameters of the Bursty and Remnant Leakers with the stellar mass-selected control samples from the Non-Leaker population. While the distribution of ionization parameters is very consistent between Remnant Leakers and Non-Leakers, we find a significant enhancement in the dimensional ionization parameter ($q$) for the Bursty Leakers compared to the other two populations. This reflects the strong enhancement in ionizing luminosity observed in Figure~\ref{pop_hists} and is certainly due to the much younger stellar populations in the Bursty Leakers as can be seen in Figure~\ref{SFH_comp} as this is the primary source if ionizing radiation in the galaxies. 

If we compare H{\small II} region gas density, we find that the Bursty Leakers exhibit similar gas densities to the Non-Leaker control sample, perhaps lacking some of the densest gas, although this may be due to limited sample size. This indicates that the ISM in the Bursty Leakers is not fully disrupted, but the combination of early SNe and radiation feedback must be clearing channels in the ISM. The fact that the Bursty leakers have typical gas densities also further demonstrates that the enhancement in ionization parameter is due to an enhancement in radiation and not a reduction in density. In contrast, the Remnant Leakers exhibit significantly lower ISM gas densities compared to the Non-Leaker control sample or the Bursty Leakers, indicating that their ISM is nearly fully disrupted. Returning to the \cite{Zackrisson2013} models, the ISM of Bursty Leakers is more akin to an ionization-bounded nebula with holes while the ISM of Remnant Leakers seem to be more representative of a density-bounded nebula.

We further demonstrate the differences in gas distribution between Bursty and Remnant Leakers in Figure~\ref{leaker_maps} where we show H{\small I} column density maps for 18 Bursty Leakers (top rows) and 18 Remnant Leakers (bottom rows). The Remnant Leaker maps have been selected by stellar mass to closely match those of the Bursty Leakers that are shown. Each image is 10~physical kpc in width. It is clear that galaxies in both leaker categories exhibit a wide diversity in H{\small I} morphology and in all systems, there is evidence for disruption in the ISM. Nevertheless, among the Bursty Leaker population, there seem to be substantially more systems with dense central clouds of neutral gas compared to the Remnant Leaker population. In contrast, the Remnant Leaker galaxy population often exhibits large cavities of ionized gas that can only be created by substantial SNe feedback. Qualitatively, there are morphological differences between the neutral gas distributions in each class of leaker. 

In Figure~\ref{leaker_time} we show a time series of the H{\small I} distribution for two example Bursty Leakers in the 85~Myr prior to $z=4.64$. In both cases, the galaxy is in an idle state until a strong burst of star formation occurs. For the first galaxy in Figure~\ref{leaker_time}, we see no evidence of any strong dynamical interactions that is driving the burst while for the second galaxy, a small gaseous clump can be seen migrating to the center, which may help drive the starburst. It is clear that the feedback from star formation is driving the increase in $f_{\rm esc}$ in both galaxies. Such behaviour was also reported in \citep{Trebitsch2017,Kimm2017,Rosdahl2018}.

\subsection{Implications for [C{\small II}] 158$\mu$m Emission}
From a theoretical viewpoint, differentiating the mechanisms by which LyC photons escape galaxies can improve our understanding of the epoch of reionization and the sources responsible. Testing the predictions from our simulations against observations is a necessary step to determine the reliability of our model. Identifying the observational consequences of each leakage mode is thus key for testing such predictions. As the [C{\small II}] 158$\mu$m emission line is one of the brightest emission lines at high-redshift \citep[e.g][]{Carilli2013}, we focus this Section on the implications of different LyC leakage modes on [C{\small II}] emission. We reiterate that it is already well established that the ratio of [O{\small III}]~88$\mu$m to [C{\small II}] emission positively correlates with $f_{\rm esc}$ \citep[e.g.][]{Inoue2016,Katz2020,Katz2021b}; however, this ratio does not describe how $f_{\rm esc}$ relates only to [C{\small II}], which is particularly interesting at $z<6$ where [OIII] is more difficult to observe with ALMA due to it falling in Band~9 or Band~10 that has a limited observational window.

\begin{figure}
\centerline{\includegraphics[scale=1,trim={0 0.0cm 0cm 0cm},clip]{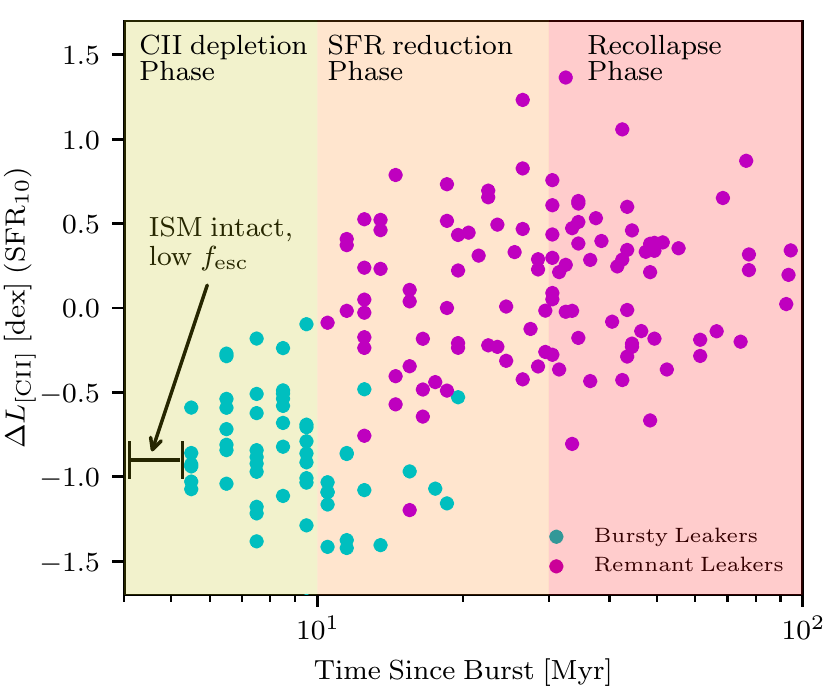}}
\centerline{\includegraphics[scale=1,trim={0 0.0cm 0cm 0cm},clip]{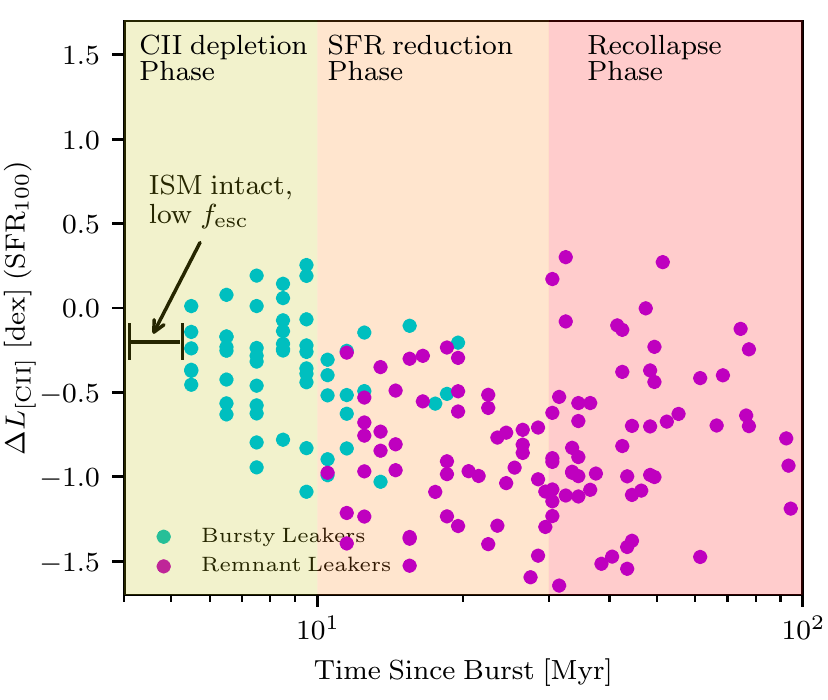}}
\caption{Deviation of the [C{\small II}] luminosity from the mean value on the [C{\small II}]-SFR$_{10}$ relation (top) or [C{\small II}]-SFR$_{100}$ relation (bottom) as a function of time since the peak SFR for SPHINX$^{20}$ leakers. Bursty and Remnant Leakers are shown in cyan and magenta, respectively. No leakers populate the first 5~Myr of the plot as it takes time for feedback to disrupt the ISM of the star forming clouds. In the top panel, galaxies initially show deficits due to radiation feedback and early SNe feedback destroying PDRs. We propose that once the feedback becomes effective enough, the SFR drops, which moves galaxies closer to the [C{\small II}]-SFR$_{10}$ relation. Once star formation is completely shut down, galaxies appear to have a [C{\small II}] excess due to residual neutral gas. Finally galaxies approach the relation as the ISM reforms and star formation begins again. In the bottom panel, the [C{\small II}] deficit continues to increase with time because the SFR is fixed over the 100~Myr interval, but SNe continue to explode over 50~Myr reducing the neutral gas content.}
\label{cii_burst}
\end{figure}

In Figure~\ref{line_sfr} we show the [C{\small II}]-SFR relation for SPHINX$^{20}$ galaxies at $z=4.64$ compared with low-redshift observational estimates from \cite{DeLooze2014}, high-redshift observational estimates from the ALPINE [C{\small II}] survey at $z\approx4.4-5.9$ \citep{Bethermin2020, Schaerer2020, Faisst2016} where only $z<5.5$ galaxies from ALPINE are shown, and the high-redshift models from \cite{Lagache2018}. The left panel shows the relation using SFR$_{10}$ while the right panel uses SFR$_{100}$. We find very little difference in the mean [C{\small II}]-SFR relation (shown as dashed red lines) regardless of which time period the SFR is measured over. At ${\rm SFR>1~M_{\odot}yr^{-1}}$ we find very good agreement between SPHINX$^{20}$ and both low- and high-redshift observations. However, the situation is fundamentally different at lower SFR. 

We find a kink in the relation at ${\rm SFR_{10,100}\sim0.1\ M_{\odot}yr^{-1}}$ such that the typical SPHINX$^{20}$ galaxy exhibits much higher [C{\small II}] luminosities at lower SFR compared to the extrapolations of the observed relations. The kink in the [C{\small II}]-SFR relation is not unique to our simulation as it is also seen in \cite{Pallottini2022}. 

Depending on whether we use SFR$_{10}$ or SFR$_{100}$ for the [C{\small II}]-SFR relation, the Bursty Leakers and Remnant Leakers populate different portions of the diagram. Beginning with SFR$_{10}$, the Bursty Leakers exhibit large [C{\small II}] deficits for their given SFR while Remnant Leakers tend to fall on the relation. Switching SFR$_{10}$ for SFR$_{100}$ (as shown in the right panel of Figure~\ref{line_sfr}), both Bursty and Remnant Leakers exhibit deficits in the [C{\small II}]-SFR relation, with the Remnant Leakers exhibiting slightly larger deficits. 

We emphasize that both the emission line signatures and SFRs of individual galaxies are dynamic and highly evolving with time \citep[e.g.][]{Barrow2020}. In fact, we view the evolution of an individual galaxy on the [C{\small II}]-SFR relation as a potential probe of state of the ISM in concert with star formation. This is highlighted in the top panel of Figure~\ref{cii_burst} where we plot the time elapsed in Myr since the burst of star formation\footnote{The time of the burst is measured at the time of peak star formation rate within the 100~Myr prior to $z=4.64$.} that caused LyC photons to leak against the [C{\small II}]-deficit (defined as the difference between the [C{\small II}] emission of a galaxy and the mean [C{\small II}] emission of all SPHINX$^{20}$ galaxies at a fixed SFR) for the [C{\small II}]-SFR$_{10}$ relation for both Bursty and Remnant Leakers. We propose the following illustrative model. For the first $4-5~{\rm Myr}$ after the burst, $f_{\rm esc}$ is low as the low column density channels have yet to form. Hence the ISM remains intact and we have no leakers in this regime. However, once enough stars form, the ionizing radiation can create holes in the ISM and reduce the neutral gas content. We describe this phase as the C{\small II} depletion phase\footnote{Note that there is a prior phase that we call the Burst phase where the SFR is increasing. This also causes [C{\small II}] deficits as discussed earlier. It is however not shown on the plot because we plot time since the maximum SFR (i.e. the period after the initial Burst phase) where the instantaneous SFR begins decreasing again. One must keep in mind that SFR$_{10}$ can be long enough to average over the most of the burst (i.e. both part of the increase and decrease, see Figure~\ref{SFH_comp}) so it takes slightly longer for SFR$_{10}$ to decrease compared to the instantaneous SFR (i.e. that averaged over 1~Myr).} and it is qualitatively shown in yellow in Figure~\ref{cii_burst}. Once the feedback becomes efficient, the SFR begins to decrease, which moves the galaxy to the left on the [C{\small II}]-SFR$_{10}$ relation, back towards the mean relation, and has the effect of reducing the [C{\small II}] deficit. At the same time SNe are exploding, which can further reduce the neutral gas content and can balance the reduction in [C{\small II}] deficit caused by reducing the SFR. Once star formation has been substantially reduced (i.e. by a few orders of magnitude), the galaxies may overshoot the [C{\small II}]-SFR relation as some neutral gas likely persists. The galaxies will loiter in this SFR reduction phase (shown in orange on Figure~\ref{cii_burst}) until the gas in the CGM can recollapse and inflows from the IGM bring fresh gas into the system. As the galaxies cool down, the ISM will begin to reform and approach the mean relation as both the SFR and [C{\small II}] emission begin to increase again (shown in red on Figure~\ref{cii_burst} as the Recollapse phase). 

The behaviour is fundamentally different when following the same evolution using the deficit on the [C{\small II}]-SFR$_{100}$ relation. In this case, the SFR for both Bursty and Remnant Leakers is enhanced compared to the typical galaxy of the same stellar mass. This moves the galaxies to the right on the [C{\small II}]-SFR$_{100}$ relation, resulting in a [C{\small II}] deficit. As the ionizing photons destroy the neutral gas content, the [C{\small II}] deficit increases. As SNe explode, the deficit continues to increase, which is why the Remnant Leakers continue the trend of increasing [C{\small II}] deficit with time. The SNe explode for $\sim50$~Myr and as this process subsides, the galaxies can begin to recollapse. Thus, it is important to note that the total C{\small II} content of a galaxy will decrease throughout both the C{\small II} depletion and SFR reduction phases, even if the galaxy moves closer to the [C{\small II}]-SFR$_{10}$ relation during this phase.

The qualitative evolution that we have described is subject to significant scatter due to various galaxy properties; hence, the [C{\small II}] deficit is not a perfect one-to-one relation with the time since the burst. For example, the length and strength of the bursts will play a role in how quickly the photodissociation regions (PDRs) are destroyed and star formation is reduced. The initial state and structure of the ISM as well as the local efficiency of star formation will also help determine how efficient the feedback is and how quickly it impacts [C{\small II}] and subsequent star formation \citep{Kimm2019,Kim2019,Kimm2022}. The strength of galactic inflows and the cooling rate in the CGM will impact the length of the loitering/SFR reduction phase as well as how quickly the gas can recollapse. There is already a substantial amount of scatter in the [C{\small II}]-SFR relation that encapsulates many of these galaxy formation processes.

We conclude this Section by noting that the idea that strong bursts of star formation lead to [C{\small II}] deficit is not new. Since the early discovery of potential [C{\small II}] deficits \citep[e.g.][]{Maiolino2015} and the latter confirmation of some \citep[e.g.][]{Laporte2019,Carniani2020}, various explanations have been proposed for this behaviour. These include the impact of the CMB and low metallicity \citep[e.g.][]{Vallini2015,Pallottini2017}, radiation field intensity \citep[e.g.][]{Lagache2018}, and bursty star formation \citep[e.g.][]{Ferrara2019,Pallottini2022}. We confirm that on the [C{\small II}]-SFR$_{10}$ relation, young bursty leakage leads to [C{\small II}] deficits while older starbursts that lead to LyC leakage typically result in normal or even enhanced [C{\small II}] for a given SFR. In all cases, strong bursts lead to deficits on the [C{\small II}]-SFR$_{100}$ relation, consistent with other work in the literature \citep[e.g.][]{Ferrara2019,Pallottini2022}.

\begin{figure*}
\centerline{\huge {$z=6$}}
\centerline{
\includegraphics[scale=1,trim={0 0.0cm 0cm 0cm},clip]{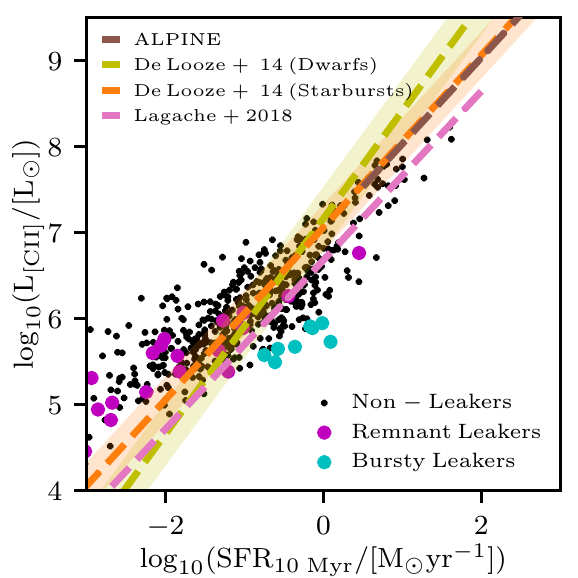}
\includegraphics[scale=1,trim={0 0.0cm 0cm 0cm},clip]{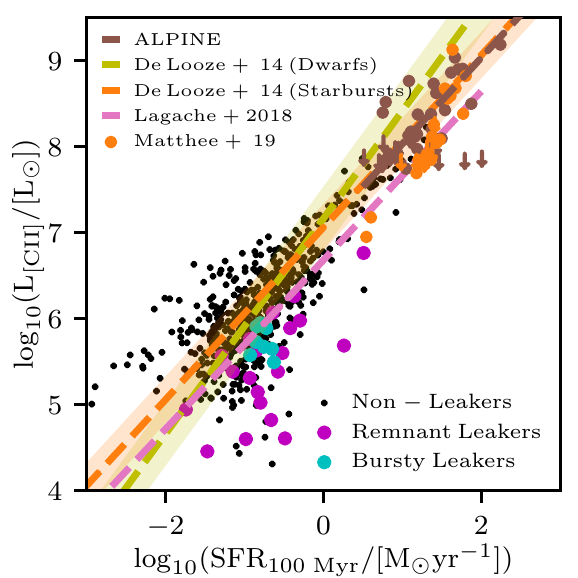}
\includegraphics[scale=1,trim={0 0.0cm 0cm 0cm},clip]{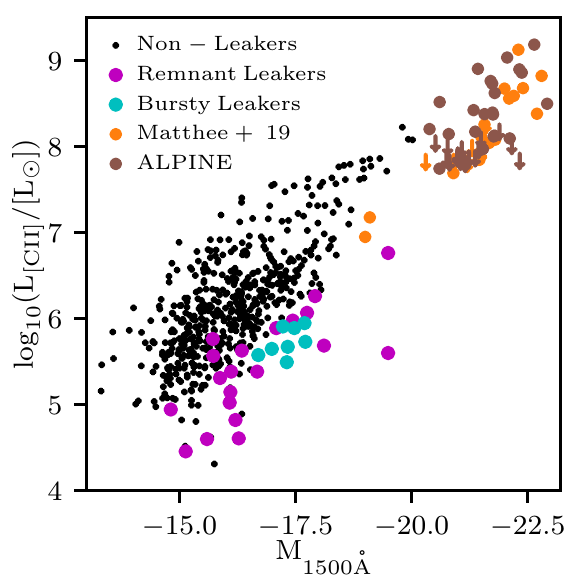}
}
\centerline{

\includegraphics[scale=1,trim={0 0.0cm 0cm 0cm},clip]{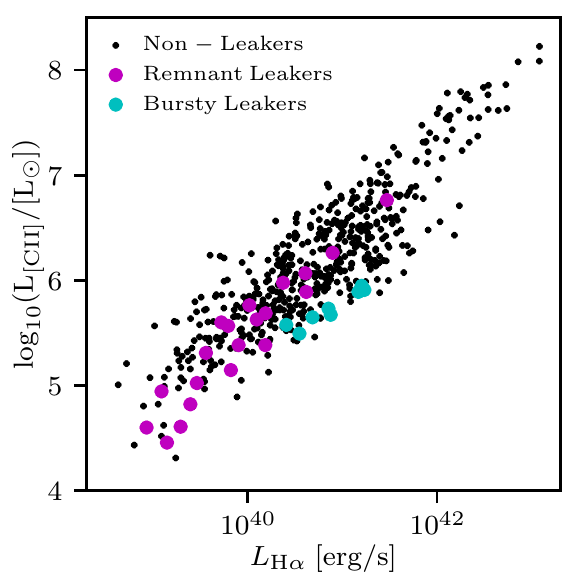}
\includegraphics[scale=1,trim={0 0.0cm 0cm 0cm},clip]{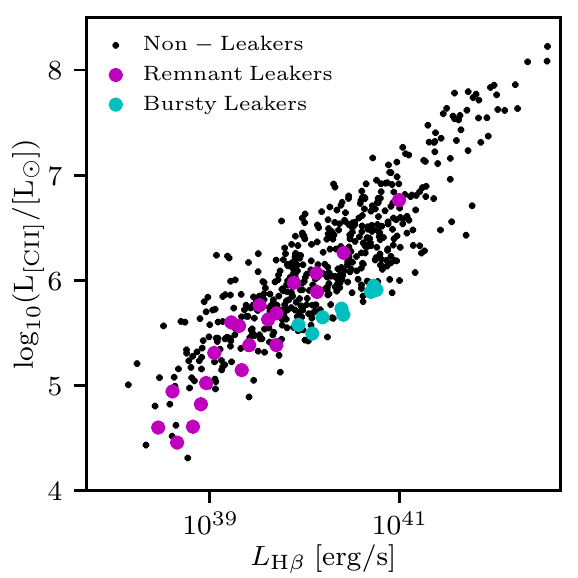}

}
\caption{[C{\small II}]-SFR$_{10}$ (top left), [C{\small II}]-SFR$_{100}$ (top centre), [C{\small II}]-1500\angstrom~UV magnitude (top right), [C{\small II}]-H$\alpha$ (bottom left), and [C{\small II}]-H$\beta$ (bottom right) relations for {\small SPHINX$^{20}$} galaxies at $z=6$ compared with fits to various types of low-redshift galaxies \protect\citep[olive-green and orange,][]{DeLooze2014} as well as high-redshift observations \protect\citep[brown,][]{Schaerer2020} and models \protect\citep[pink,][]{Lagache2018}. Bursty Leakers are shown in cyan while Remnant Leakers are shown in magenta. Non-Leakers are shown in black. Galaxies are only included if their halo mass is $>3\times10^9{\rm M_{\odot}}$. Individual $z>5.5$ ALPINE galaxies are shown as brown points for detections or arrows representing $3\sigma$ upper limits on [C{\small II}] luminosity. Orange points represent the high-redshift data compilation from \protect\cite{Matthee2019b}.}
\label{line_sfr_z6}
\end{figure*}

\subsection{Finding LyC Leakers in and out of the Epoch of Reionization with [C{\small II}]}
We have shown that the location of a LyC leaker on both the [C{\small II}]-SFR$_{10}$ and [C{\small II}]-SFR$_{100}$ relations can differentiate the type of LyC leakage that is occurring in the galaxy. The primary motivation for studying this physics at $z<6$ is because IGM attenuation prevents direct observations of LyC emission during the epoch of reionization (EoR). Furthermore, our predictions can be tested at lower redshift with known LyC leakers. However, as our ultimate goal is to understand LyC leakage in the EoR, we continue our analysis by showing that our results at $z=4.64$ also hold at $z=6$ and that current large ALMA programs \citep[e.g.][]{LeFevre2020,Bouwens2021a} focusing on [C{\small II}] emitters at high-redshift may be able to detect LyC leakers. In order to best compare with observations, we focus our analysis on the $\sim700$ most massive haloes (i.e. haloes with virial masses $>3\times10^{9}{\rm M_{\odot}}$) in SPHINX$^{20}$ at $z=6$. In this sample, we find 8 Bursty Leakers, 20 Remnant Leakers, and 515 Non-Leakers.  The distribution of Bursty to Remnant leakers is similar to the $z=4.64$ snapshot. 

In the top left and top centre panels of Figure~\ref{line_sfr_z6} we show the [C{\small II}]-SFR$_{10}$ and [C{\small II}]-SFR$_{100}$ relations for the massive SPHINX$^{20}$ galaxies at $z=6$. The behaviour is identical to that seen in Figure~\ref{line_sfr} for $z=4.64$. When using SFR$_{10}$, Bursty Leakers show [C{\small II}] deficits, while Remnant Leakers populate similar regions as Non-leakers. In contrast, when using SFR$_{100}$ both types of leaker exhibit [C{\small II}] deficits. For comparison with observations, the top centre panel also contains detections as well as $3\sigma$ limits on [C{\small II}] for spectroscopically confirmed galaxies at $z>5.5$ from the ALPINE survey and at $z>6$ from the compilation by \citet{Matthee2019b}. 

In the bottom left and bottom right panels of Figure~\ref{line_sfr_z6}, we show the [C{\small II}]-H$\alpha$\footnote{Here we use the intrinsic H$\alpha$ emission rather than the dust attenuated value. This assumes that the dust content of the galaxy can be inferred and attenuation corrected for. We do the same for H$\beta$.} and [C{\small II}]-H$\beta$ relations, respectively. We find a strong linear trend between the log values of [C{\small II}] and both Balmer emission lines, which reflects the fact that both are considered SFR indicators \citep[e.g.][]{Kennicutt1992,DeLooze2014}, despite the fact that they probe different gas (i.e. neutral versus ionized). Bursty leakers once again exhibit [C{\small II}] deficits while remnant leakers are more consistent with the bulk of the SPHINX$^{20}$ galaxy population at $z=6$. To probe longer time scale star formation, in the top right panel of Figure~\ref{line_sfr_z6}, we compare [C{\small II}] emission with the 1500\angstrom~UV magnitude\footnote{Here we are using the angle-averaged dust attenuated value for UV magnitude. The results are the same if we use the intrinsic magnitude rather than the dust attenuated value.} of the galaxy. We now see that both Bursty and Remnant Leakers exhibit [C{\small II}] deficits compared to other SPHINX$^{20}$ galaxies at the same UV magnitude. Thus the combination of [C{\small II}] luminosity, UV magnitude, and H$\alpha$ and H$\beta$ emission are very powerful for identifying Bursty and Remnant leakers, and potentially differentiating the two.

However, we highlight that there are potential caveats with this approach. The observed LyC escape fraction is highly viewing angle-dependent \citep[e.g.][]{Cen2015}. Not all galaxies with [C{\small II}] deficits will be observed as leakers, both because there are some galaxies with deficits that are truly Non-Leakers (i.e. the [C{\small II}] deficit galaxies are not purely comprised of leakers), and due to the viewing angle dependence, an observer may be unlucky and be positioned along an optically thick line of sight. Only large samples of galaxies will be able to disentangle this degeneracy and this is further discussed below. Thus to confirm our predictions, we recommend initially studying known LyC leakers and investigating whether they exhibit the relevant [C{\small II}] deficits. However, this also introduces biases as the population of known LyC leakers is subject to various selection effects; nevertheless, we expect [C{\small II}] deficits should persist, even for biased samples of leakers, when using a SFR indicator sensitive to long time scales. We have also assumed that the intrinsic H$\alpha$ and H$\beta$ emission can be derived with the appropriate dust corrections while the angle-averaged 1500\angstrom~attenuation is representative of what one would observe along a typical line of sight. Viewing angle effects can also introduce additional scatter into the relations. Since [C{\small II}] emission originates primarily in neutral gas at these redshifts \citep{Pallottini2017,Katz2019,Lupi2020}, H$\alpha$ and H$\beta$ emission comes from mostly ionized (or partially ionized) gas, and LyC photons and 1500\angstrom~photons come from stars, differences in attenuation for each of these sources along various lines of sight can be important. The simplistic dust modelling employed in SPHINX$^{20}$ in post-processing can account for this \citep{Katz2022-mgII}; however, future simulations with more self-consistent modelling will be required to further assess this additional scatter. 

\section{Discussion \& Conclusions}
\label{conclusions}

\begin{figure}
\centerline{
\includegraphics[scale=1,trim={0 0.0cm 0cm 0cm},clip]{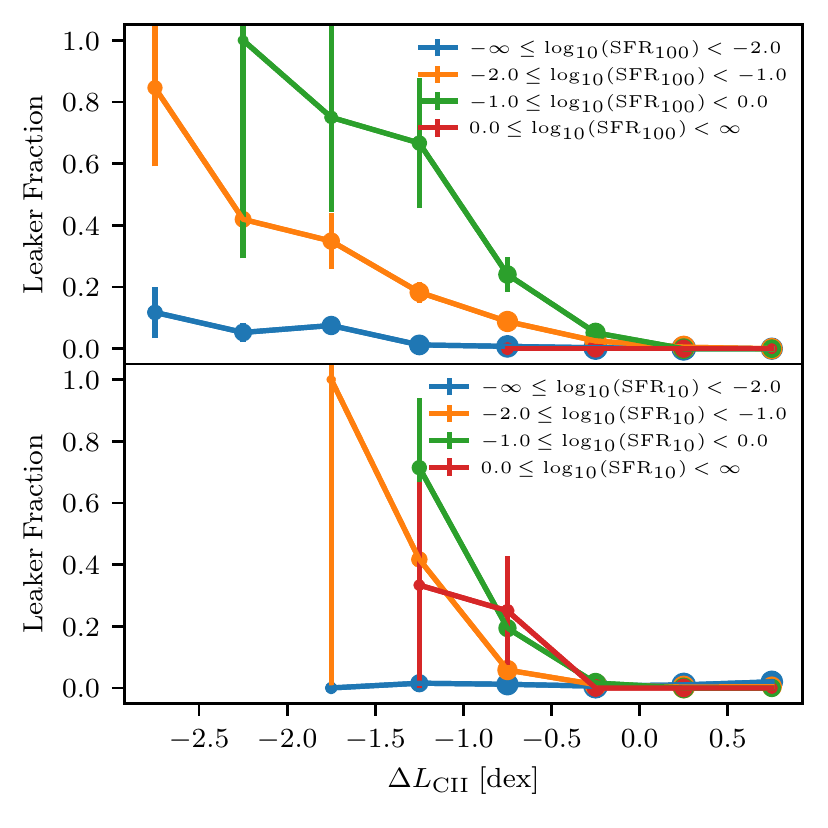}
}
\caption{Fraction of galaxies that have $f_{\rm esc}>20\%$ as a function of deviation from the [C{\small II}]-SFR relation for different bins of star formation rate. The top and bottom panels show the results using SFR$_{100}$ and SFR$_{10}$, respectively.}
\label{leak_frac}
\end{figure}

\subsection{Recommended Observing Strategy}
Our results excite great optimism that [C{\small II}] can be combined with other commonly observed quantities (e.g. UV magnitude or H$\alpha$ luminosity) to not only identify potential LyC leakers, but also gauge the mechanisms by which the LyC radiation is leaking. However, we emphasize that our results represent population averages and results for individual haloes may vary due to numerous observational (e.g. orientation angle) and physical effects. For this reason, we argue that the best way to test our model is to follow up known low-redshift LyC leakers with [C{\small II}] observations. This will remove issues related to line of sight effects, as all leakers (i.e. regardless of being Bursty or Remnant) must show [C{\small II}] deficits at some level when using a SFR indicator sensitive to $\sim100$~Myr time scales.

ALMA is sensitive to [C{\small II}] in Band~8 at $3\lesssim z \lesssim 4$. The relatively restrictive transmission function of ALMA Band~8 means that [C{\small II}] observations from galaxies in only certain redshift bands are possible. Nevertheless, with the increasingly large samples of known LyC leakers in this redshift interval \citep[see][for a recent compilation]{Mestric21}, there are enough galaxies to begin testing the predictions made in this work.

We highlight a few important considerations to keep in mind when testing our predictions. We note that not all galaxies with [C{\small II}] deficits are LyC leakers. In Figure~\ref{leak_frac} we show the probability of a galaxy being a LyC leaker as a function of [C{\small II}] deficit for both the [C{\small II}]-SFR$_{10}$ and [C{\small II}]-SFR$_{100}$ relations binned by SFR. It is clear that this fraction varies significantly between SFR bins and as a function of [C{\small II}] deficit. On the [C{\small II}]-SFR$_{100}$ relation, a [C{\small II}] deficit seems to be a necessary, but insufficient condition for being a LyC leaker. Furthermore, the direct detection of LyC leakage suffers from both orientation angle effects and IGM transmissivity. This means that even if a galaxy exhibits high [C{\small II}] deficits, significant LyC leakage may not be detectable for a particular galaxy. Finally, when following up known LyC leakers with [C{\small II}] observations, it is also important to consider orientation angle effects as well because an individual sight line in a galaxy may be optically thin even if the angle-averaged LyC $f_{\rm esc}$ is low, in which case no [C{\small II}] deficit may be present. For example, we find that the number of lines of sight with $f_{\rm esc}>20\%$ that are hosted by galaxies that have angle-averaged $f_{\rm esc}>20\%$ is a ratio of 2 to 1 (see \citealt{Katz2022-mgII}). For these reasons, large samples of galaxies will be needed to overcome the scatter introduced by these biases.

\begin{figure}
\centerline{
\includegraphics[scale=1,trim={0 0.0cm 0cm 0cm},clip]{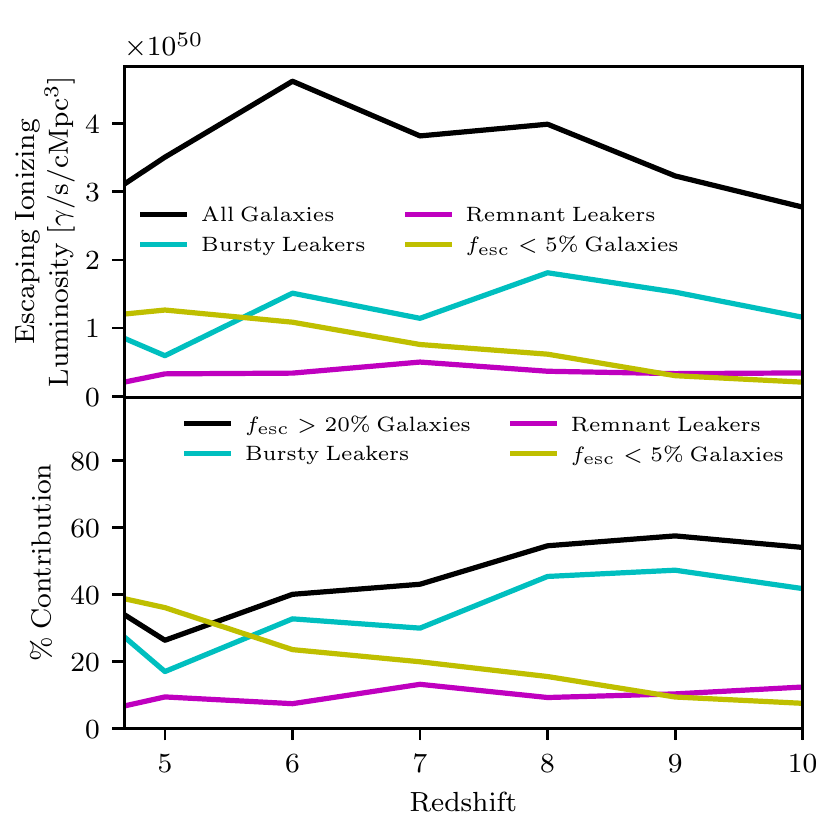}
}
\caption{(Top) Ionizing luminosity escaping into the IGM as a function of redshift in SPHINX$^{20}$. The black line shows the contribution from all galaxies while the cyan and magenta lines show the total escaping ionizing luminosities of Bursty and Remnant Leakers, respectively. The olive-green line shows the contribution from galaxies with $f_{\rm esc}<5\%$. (Bottom) Percentage contribution to the total escaping ionizing budget for Bursty and Remnant Leakers as a function of redshift.  }
\label{leak_cont}
\end{figure}

\subsection{Contribution to the Reionization Budget}
We have focused primarily on two types of LyC leakers that were arbitrarily defined to have LyC escape fractions $>20\%$. An open question is what role do these systems play in the reionization of the Universe and whether it changes between $z\approx4.64$ and $z\approx6$. To understand this, we quantify the contribution of each type of leaker to the reionization process as a function of redshift. In Figure~\ref{leak_cont} we show the total luminosity of ionizing photons escaping into the IGM as a function of redshift compared to the contribution from Bursty and Remnant Leakers. 

At all redshifts, Remnant Leakers represent a subdominant contribution to the total escaping ionizing luminosity budget. Despite their higher number densities, the lack of recent star formation means they have much lower intrinsic ionizing luminosities compared to Bursty Leakers. Although subdominant, the Remnant Leaker photon budget is non-negligible and represents up to $\sim10\%$ of the total at any redshift. Bursty Leakers contribute significantly more ionizing photons, contributing up to $\sim47\%$ at high-redshift. The relative contribution from both Bursty and Remnant Leakers decreases with decreasing redshift. 

We interpret this to mean that galaxies with high $f_{\rm esc}$ are a necessary and important part of reionization, but low $f_{\rm esc}$ galaxies are equally important for reionization. To demonstrate this, we show the contribution of galaxies with $f_{\rm esc}<5\%$ to the reionization budget as olive-green lines. At the end of reionization, the low $f_{\rm esc}$ galaxies contribute equal numbers of ionizing photons into the IGM as the high $f_{\rm esc}$ galaxies. It should be noted that this does not necessarily imply that there are numerous galaxies with low $f_{\rm esc}$ all contributing equally. At any given redshift, the reionization budget is dominated by a small fraction of halos. Since intrinsic ionizing luminosities increase with decreasing redshift (as galaxies become more massive), low $f_{\rm esc}$ galaxies can still provide substantial numbers of ionizing photons to the IGM. This is further discussed in \cite{Rosdahl2022}. However, a potential caveat to these conclusions is that SPHINX$^{20}$ reionizes slightly later than the reionization history inferred from observations \citep[e.g.][]{Fan2006,Kulkarni2019}. A detailed discussion into the numerics of this is beyond the scope of this paper; however, different solutions will have varying impacts on Figure~\ref{leak_cont} and this warrants further exploration. 

\subsection{Caveats}
As our work is based on cosmological radiation hydrodynamics simulations, various caveats should be considered when interpreting our results. A detailed discussion of these can be found in \cite{Katz2021b} and we summarize the primary uncertainties of our model below.

Like all simulations, SPHINX$^{20}$ has finite spatial and mass resolution and thus quantities such as the LyC escape fraction can only be measured on the scale (i.e. 10~pc) at which they are injected into the simulation. Important ISM physics may be occurring at scales that are not resolved by the simulation. SPHINX$^{20}$ includes various sub-grid models for star formation and feedback. Such models are designed to reasonably reproduce the behaviour of this physics on the scales that are resolved, but similar luminosity functions or stellar mass-halo mass relations may be obtained with another choice of model. These choices can all impact [C{\small II}] emission, SFRs, and LyC escape. Furthermore, due to the volume of SPHINX$^{20}$, the stellar mass range probed by the simulation does not yet overlap with observed high-redshift galaxy samples (e.g. ALPINE or REBELS). SPHINX$^{20}$ predicts that high mass galaxies have very low $f_{\rm esc}$ in general, so observations may need to push to lower stellar masses for our predictions to be robustly tested.

Extracting [C{\small II}] emission from simulations is highly non-trivial (see e.g. \citealt{Olsen2018}). Because SPHINX$^{20}$ does not compute the non-equilibrium C{\small II} abundance or level populations self-consistently, such measurements must be made in post-processing. While we have attempted to use as much of the information as possible from the simulation in post-processing, there are multiple methods for extracting [C{\small II}] luminosity that may result in different values. Our results are in good agreement with observational constraints, especially at high stellar masses, which should provide confidence in our method; however future simulations \citep[e.g.][]{Katz2022a} that follow the non-equilibrium physics in a more self-consistent manner will be needed to confirm our results. 

\subsection{Summary}
We have analyzed the SPHINX$^{20}$ simulation at $z=4.64$ and $z=6$ to understand the observational signatures of escaping LyC photons. Focusing primarily on [C{\small II}]$_{\rm 158\mu m}$ emission, we identified two classes of galaxies with high $f_{\rm esc}$ (i.e. $f_{\rm esc}>20\%$): Bursty Leakers and Remnant Leakers. The former are categorized as having a very recent burst of star formation within the past 5-15 Myr, exceptionally high ionization parameters and typical gas densities. The latter are remnants of a much older burst of star formation (within the past 100~Myr) and exhibit very little current star formation, low ISM gas densities, and normal ionization parameters. The two classes of leakers are qualitatively similar to ionization bounded nebulae with holes and density bounded nebula, respectively. We compared and contrasted each type of leaker with each other and the Non-Leaker population and our conclusions can be summarised as follows:
\begin{itemize}
    \item To reach a LyC escape fraction of $>20\%$, galaxies typically undergo bursts of star formation that are $90\times-100\times$ higher (when averaged in 1~Myr intervals) than the SFR of the galaxy prior to the burst. Not all bursts of star formation of such magnitude are guaranteed to result in efficient LyC leakage. The primary difference between whether a galaxy is a Bursty Leaker or a Remnant Leaker is the timing of the burst of star formation.
    \item When compared with Non-Leakers on the galaxy formation main sequence (${\rm SFR-M_*}$), both Bursty and Remnant Leakers populate the regions above the mean when using an SFR indicator that is sensitive to timescales of $\sim100$~Myr, which is consistent with observations. For shorter time scale indicators (e.g. H$\alpha$), only the Bursty Leakers appear as outliers above the mean.
    \item When using short time scale SFR indicators (e.g. H$\alpha$), only Bursty Leakers show [C{\small II}] deficits. In contrast, when using a longer time scale indicator (e.g. UV magnitude), both types of leaker show [C{\small II}] deficits. The combination of observations of [C{\small II}] as well as short (e.g. H$\alpha$ and H$\beta$) and long time scale SFR indicators (e.g. M$_{\rm UV}$) is a powerful combination for not only identifying potential LyC leakers, but also to constrain the mode by which the photons are leaking.
    \item Although [C{\small II}] deficits are promising tracers of significant LyC escape, we note that several considerations must be made when inferring [C {\small II}] deficits from LyC leakers and vice versa. Importantly, the fraction of galaxies with high [C{\small II}] deficits that also show strong LyC leakage is dependent on the SFR. Orientation and line-of-sight effects can impact the measurement of LyC leakage but are unlikely to affect [C{\small II}] luminosities. Finally, although an observed LyC leaking galaxy may exhibit a high escape fraction along the line-of-sight aligned with the observer, its angle-averaged LyC escape fraction may still be low, potentially weakening the [C{\small II}] deficit observed.
    \item Despite having LyC escape fractions of $>20\%$, the total contribution of Bursty and Remnant Leakers only peaks at $\sim60\%$ at $z=9$ and decreases thereafter. At $z\lesssim5.5$, galaxies with $f_{\rm esc}<5\%$ contribute more ionizing photons to the UV background than galaxies with $f_{\rm esc}>20\%$. Thus, it is the high $f_{\rm esc}$ galaxies that dominate early reionization, but low $f_{\rm esc}$ galaxies that complete and maintain it.
\end{itemize}

The ideas presented in this work can be tested with a reasonable set of [C{\small II}] observations in the intermediate-redshift Universe. Constraining models for LyC leakage is paramount for understanding the reionization era and galaxy formation at $z\geq6$. Thus connecting observations and simulations in this regard should be considered a top priority, especially in the context of upcoming JWST observations.

\section*{Acknowledgements}
We thank the referee for their comments which improved the manuscript. AS and RSE acknowledge financial support from European Research Council Advanced Grant FP7/669253. TG is supported by the ERC Starting grant 757258 ‘TRIPLE’. TK was supported by the National Research Foundation of Korea (NRF) grant funded by the Korea government (No. 2020R1C1C1007079 and No. 2022R1A6A1A03053472). Computing time for this work was provided by the Partnership for Advanced Computing in Europe (PRACE) as part of the “First luminous objects and reionization with SPHINX (cont.)” (2016153539, 2018184362, 2019215124) project. We thank Philipp Otte and Filipe Guimaraes for helpful support throughout the project and for the extra storage they provided us. We also thank GENCI for providing additional computing resources under GENCI grant A0070410560. This project has received funding from the European Research Council (ERC) under the European Union’s Horizon 2020 research and innovation programme (grant agreement No 693024). Some of this work used the DiRAC@Durham facility managed by the Institute for Computational Cosmology on behalf of the STFC DiRAC HPC Facility (www.dirac.ac.uk). The equipment was funded by BEIS capital funding via STFC capital grants ST/P002293/1, ST/R002371/1 and ST/S002502/1, Durham University and STFC operations grant ST/R000832/1. Some of this work was performed using the DiRAC Data Intensive service at Leicester, operated by the University of Leicester IT Services, which forms part of the STFC DiRAC HPC Facility (www.dirac.ac.uk). The equipment was funded by BEIS capital funding via STFC capital grants ST/K000373/1 and ST/R002363/1 and STFC DiRAC Operations grant ST/R001014/1. DiRAC is part of the National e-Infrastructure.

%%%%%%%%%%%%%%%%%%%%%%%%%%%%%%%%%%%%%%%%%%%%%%%%%%
\section*{Data Availability}
The data underlying this article will be shared on reasonable request to the corresponding author.

%%%%%%%%%%%%%%%%%%%% REFERENCES %%%%%%%%%%%%%%%%%%
\bibliographystyle{mnras}
\bibliography{example} 

% Don't change these lines
\bsp	% typesetting comment
\label{lastpage}
\end{document}